\documentclass[twocolumn,10pt]{article}
\usepackage{usenix}
\usepackage{fullpage}
\usepackage{graphicx}
\usepackage[utf8]{inputenc}
\usepackage{subcaption}
\usepackage{xcolor}
\usepackage{amsmath}
\usepackage{amsfonts}
\PassOptionsToPackage{hyphens}{url}\usepackage{hyperref}
\usepackage{cleveref}
\usepackage{enumitem}
\usepackage{xspace}
\usepackage{overpic}
\usepackage{array}
\usepackage{booktabs}
\usepackage{multirow}

\newif\ifarxiv

\arxivtrue

\newcommand{\Ex}{\mathop{\mathbb{E}}}
\newcommand{\xgen}{x_{\textrm{gen}}}
\newcommand{\gen}{\texttt{Gen}\xspace}
\newcommand{\tightparagraph}[1]{\medskip \noindent \textbf{#1}~}
\newcommand{\errorthr}{\delta}

\iftrue
\newcommand{\jhnote}[1]{\textcolor{blue}{jamie: #1}}
\newcommand{\mcj}[1]{\textcolor{orange}{mcj: #1}}
\newcommand{\vikash}[1]{\textcolor{brown}{vikash: #1}}
\newcommand{\borja}[1]{\textcolor{green}{borja: #1}}
\newcommand{\dei}[1]{\textcolor{purple}{dei: #1}}
\newcommand{\eric}[1]{\textcolor{purple}{Eric: #1}}
\newcommand{\milad}[1]{\textcolor{cyan}{milad: #1}}
\newcommand{\ft}[1]{\textcolor{red}{FT: #1}}
\newcommand{\TODO}[1]{\textbf{\textcolor{red}{TODO: \{#1\}}}}
\fi

\title{Extracting Training Data from Diffusion Models}

\author{
Nicholas Carlini$^{*1}$ \quad 
Jamie Hayes$^{*2}$ \quad 
Milad Nasr$^{*1}$
\\
Matthew Jagielski$^{+1}$ \quad 
Vikash Sehwag$^{+4}$ \quad 
Florian Tramèr$^{+3}$
\\
Borja Balle$^{\dag2}$ \quad
Daphne Ippolito$^{\dag1}$ \quad
Eric Wallace$^{\dag5}$ \\
\emph{
$^1$Google \quad 
$^2$DeepMind \quad 
$^3$ETHZ \quad 
$^4$Princeton \quad 
$^5$UC Berkeley
} \\
\emph{$^*$Equal contribution\quad $^+$Equal contribution\quad $^\dag$Equal contribution} \\
\,
\\
}

\usepackage{letltxmacro}
\usepackage{pgffor}

\newcommand\FigList{}
\newcommand\AddFigToList[1]{\edef\FigList{\FigList#1,}}

\LetLtxMacro{\OldIncludegraphics}{\includegraphics}
\renewcommand{\includegraphics}[2][]{%
    \AddFigToList{#2}%
    \OldIncludegraphics[#1]{#2}%
}

\newcommand*{\ShowListOfFigures}{%
    \typeout{Figures included were}%
    \foreach \x in \FigList {%
        \typeout{ \x}
    }%
}
\AtEndDocument{\ShowListOfFigures}

\date{}

\begin{document}

\maketitle

\begin{abstract}
    Image diffusion models such as DALL-E 2, Imagen, and Stable Diffusion have attracted significant attention due to their ability to generate high-quality synthetic images. 
    In this work, we show that diffusion models memorize individual images from their training data and emit them at generation time. 
    With a generate-and-filter pipeline, we extract over a thousand training examples from state-of-the-art models, ranging from photographs of individual people to trademarked company logos. 
    We also train hundreds of diffusion models in various settings to analyze how different modeling and data decisions affect privacy. 
    Overall, our results show that diffusion models are much less private than prior generative models such as GANs, and that mitigating these vulnerabilities may require new advances in privacy-preserving training.
\end{abstract}

\section{Introduction}

Denoising diffusion models are an emerging class of generative neural networks that produce images from a training distribution via an iterative denoising process~\cite{sohl2015deep, song2019generative, ho2020denoising}. Compared to prior approaches such as GANs~\cite{goodfellow2020generative} or VAEs~\cite{kingma2013auto},
diffusion models produce higher-quality samples~\cite{dhariwal2021diffusion} and are easier to scale~\cite{ramesh2022hierarchical} and control~\cite{nichol2021glide}.
Consequently, they have 
rapidly become the de-facto method for generating high-resolution images, and large-scale models such as DALL-E 2~\cite{ramesh2022hierarchical}
have attracted significant public interest. 

%
The appeal of generative diffusion models is rooted in their ability to synthesize novel images that are ostensibly unlike anything in the training set. 
Indeed, past large-scale training efforts ``do not find overfitting to be an issue'', \cite{saharia2022photorealistic} 
and researchers in privacy-sensitive domains have even suggested that diffusion models could ``protect[] the privacy [...] of real images''~\cite{jahanian2021generative} by generating synthetic examples~\cite{Chambon2,chambon,Rouzrokh,Ali,pinaya}. 
This line of work relies on the assumption that diffusion models do not memorize and regenerate their training data. If they did, it would violate all privacy guarantees and raise numerous questions regarding model generalization and ``digital forgery'' \cite{somepalli2022diffusion}.

\begin{figure}
\centering
\vspace{-.1in}
\includegraphics[scale=.32]{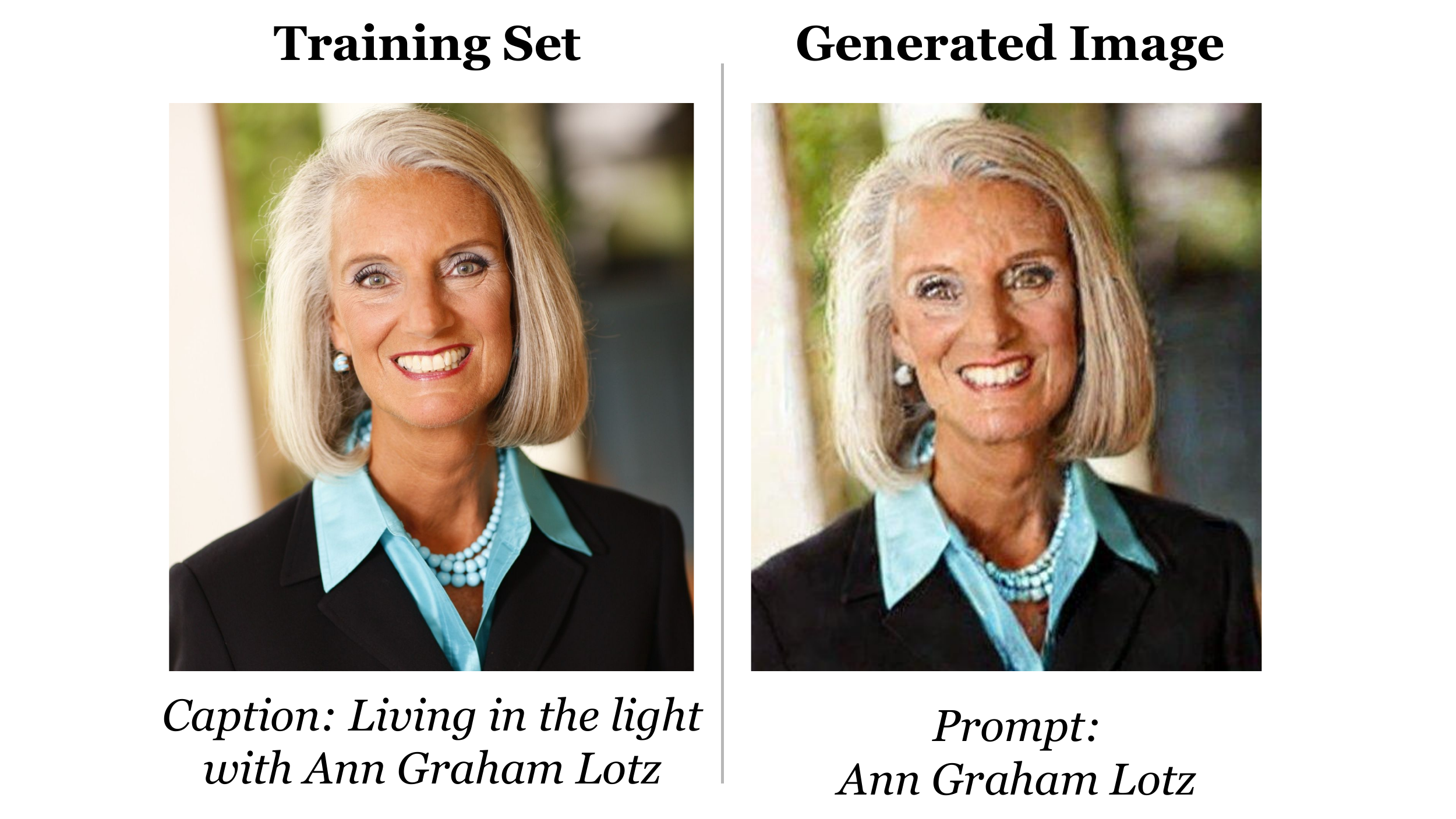}
\vspace{-0.65cm}
\caption{Diffusion models memorize individual training examples 
and generate them at test time. 
\textbf{Left:} an image from Stable Diffusion's training set (licensed CC BY-SA 3.0, see~\cite{annlotz}). \textbf{Right:} a Stable Diffusion generation when prompted with ``Ann Graham Lotz''.
The reconstruction is nearly identical ($\ell_2$ distance = 0.031).}
\label{fig:teaser}
\end{figure}


In this work, we demonstrate that state-of-the-art diffusion models \emph{do} memorize and regenerate individual training examples.
To begin, we propose and implement new definitions for ``memorization'' in image models. We then devise a two-stage data extraction attack that generates images using standard approaches, and flags those that exceed certain membership inference scoring criteria. Applying this method to Stable Diffusion~\cite{rombach2022high} and Imagen~\cite{saharia2022photorealistic}, we extract over a hundred near-identical replicas of training images that range from personally identifiable photos to trademarked logos (e.g., \Cref{fig:teaser}). 
%

To better understand how and why memorization occurs, we train hundreds of diffusion models on CIFAR-10 to analyze the impact of model accuracy, hyperparameters, augmentation, and deduplication on privacy.
Diffusion models are the least private form of image models that we evaluate---for example, they leak more than twice as much training data as GANs. 
Unfortunately, we also find that existing privacy-enhancing techniques 
do not provide an acceptable privacy-utility tradeoff. 
Overall, our paper highlights the tension between increasingly powerful generative models and data privacy, and raises questions on how diffusion models work and how they should be responsibly deployed.

\section{Background}

\textbf{Diffusion models.}
Generative image models have a long history 
(see~\cite[Chapter 20]{goodfellow2016deep}).
Generative Adversarial Networks (GANs)~\cite{goodfellow2020generative} were the breakthrough that first enabled the generation of high-fidelity images at scale~\cite{brock2018large, karras2019style}.
But over the last two years, diffusion models~\cite{sohl2015deep} have largely displaced GANs: they achieve state-of-the-art results on academic benchmarks~\cite{dhariwal2021diffusion} and form the basis of all recently popularized image generators such as Stable Diffusion~\cite{rombach2022high}, DALL-E 2~\cite{ramesh2021zero, ramesh2022hierarchical}, Runway~\cite{rombach2022high}, Midjourney \cite{midjourney} and Imagen~\cite{saharia2022photorealistic}.

\emph{Denoising Diffusion Probabilistic Models}~\cite{ho2020denoising}
\footnote{Our description of diffusion models below omits a number of significant 
details.
%
However, these details are orthogonal to the results of our attacks and we omit them for
simplicity.}
are conceptually simple: 
they are nothing more than image \emph{denoisers}.
During training, given a clean image $x$,
we sample a time-step $t \in [0,T]$ 
and a Gaussian noise vector $\epsilon \sim \mathcal{N}(0, I)$,
to produce a noised image $x' \gets \sqrt{a_t} x + \sqrt{1 - a_t} \epsilon$, for some decaying parameter $a_t \in [0,1]$ where $a_0 = 1$ and $a_T = 0$.
A diffusion model $f_{\theta}$ removes the noise $\epsilon$ to recover the
original image $x$ by predicting the noise that was added by stochastically minimizing the objective 
$\frac{1}{N} \sum_i \Ex_{t, \epsilon} \mathcal{L}(x_i, t, \epsilon; f_{\theta})$, where
\begin{align}\label{eqn:diffusion_loss}
    \mathcal{L}(x_i, t, \epsilon; f_{\theta}) = \lVert \epsilon - f_{\theta}(\sqrt{a_t} x_i + \sqrt{1 - a_t} \epsilon, t) \rVert_2^2 \enspace.
\end{align}

Despite being trained with this simple denoising objective, 
diffusion models can \emph{generate} high-quality images
by first sampling a random vector $z_T \sim \mathcal{N}(0, I)$ and then applying the diffusion model $f_\theta$ to remove the noise from this random ``image''.
%
%
To make the denoising process easier, we do not remove all of the noise at once---we instead iteratively apply the model to slowly remove noise.
Formally, the final image $z_0$ is obtained from $z_T$ by iterating the rule $z_{t-1} = f_{\theta}(z_t, t) + \sigma_t \mathcal{N}(0,I)$ for a noise schedule $\sigma_t$ (dependent on $a_t$) with $\sigma_1 = 0$. 
This process relies on the fact that the model $f_\theta$ was trained to denoise images with varying degrees of noise.
Overall, running this iterative generation process (which we will denote by $\gen$) with large-scale diffusion models produces results that resemble natural images.

Some diffusion models are further \emph{conditioned} to generate a particular type of image.
Class-conditional diffusion models take as input a class-label (e.g., ``dog'' or ``cat'')
alongside the noised image to produce a particular class of image.
Text-conditioned models take this one step further and take as input the text embedding of some \emph{prompt} (e.g., ``a photograph of a horse on the moon'')
using a pre-trained language encoder (e.g., CLIP~\cite{radford2021learning}).

\tightparagraph{Training data privacy attacks.}
Neural networks often leak details of their training datasets.
Membership inference attacks \cite{DBLP:conf/sp/ShokriSSS17,yeom2018privacy,carlini2022membership} answer the question
``was this example in the training set?'' and present a mild privacy breach.
Neural networks are also vulnerable to more powerful attacks such as inversion attacks~\cite{fredrikson2015model,zhang2020secret} that extract representative examples from a target class,
attribute inference attacks~\cite{fredrikson2014privacy} that reconstruct subsets of attributes of training examples,
and extraction attacks \cite{carlini2019secret,carlini2021extracting,balle2022reconstructing} that completely recover training examples.
In this paper, we focus on each of these three attacks when applied to diffusion models.

Concurrent work explores the privacy of diffusion models.
Wu \emph{et al.} \cite{wu2022membership} and Hu \emph{et al.} \cite{hu2023membership} perform membership inference attacks on diffusion models; our results use more sophisticated attack methods and study stronger privacy risks such as data extraction.
Somepalli \emph{et al.} \cite{somepalli2022diffusion} show several cases where (non-adversarially) sampling from a diffusion model can produce memorized
training examples. However, they focus mainly on comparing the semantic similarity of generated images to the training set, i.e., ``style copying''.
In contrast, we focus on worst-case privacy under a much more restrictive notion of memorization, and perform our attacks on a wider range of models.


\section{Motivation and Threat Model}

There are two distinct motivations for understanding how diffusion models memorize and regenerate training data.

\tightparagraph{Understanding privacy risks.}
Diffusion models 
that regenerate data scraped from the Internet can pose similar privacy and copyright risks as language models~\cite{carlini2021extracting, brown2022does, henderson2018ethical}. For example, memorizing and regenerating copyrighted text~\cite{carlini2021extracting} and source code~\cite{ippolito2022preventing} has been pointed to as indicators of potential copyright infringement~\cite{copilot_lawsuit}. Similarly, copying images from professional artists has been called ``digital forgery''~\cite{somepalli2022diffusion} and has spurred debate in the art community.

Future diffusion models might also be trained on more sensitive private data.
Indeed, GANs have already been applied to medical imagery \cite{tucker2020generating,dumont2021overcoming, kazeminia2020gans}, which underlines the importance of
understanding the risks of generative models \emph{before} we apply them to private domains.

Worse, a growing literature suggests that diffusion models could create synthetic training data to ``protect the privacy and usage rights of real images''~\cite{jahanian2021generative},
and production tools already claim to use diffusion models to protect data privacy~\cite{forbes_synthetic_data, synthetic_data_vendors, diffusion_document_synthesis}.
Our work shows diffusion models may be unfit for this purpose.

%

\tightparagraph{Understanding generalization.}
Beyond data privacy, understanding how and why diffusion models memorize training data may help us understand their generalization capabilities.
For instance, a common question for large-scale generative models is whether their impressive results arise from truly novel generations, or are instead the result of direct copying and remixing of their training data. By studying memorization, we can provide a concrete empirical characterization of the rates at which generative models perform such data copying.

In their diffusion model, Saharia \emph{et al.}
``do not find over-fitting to be an issue, and believe further training might improve overall performance`` \cite{saharia2022photorealistic}, and yet we will show that this model memorizes individual examples.
It may thus be necessary to broaden our definitions of overfitting to include memorization and related privacy metrics.
Our results also suggest that Feldman's theory that memorization is \textit{necessary} for generalization in classifiers~\cite{feldman2020does} may extend to generative models,
raising the question of whether the improved performance of diffusion models compared to prior approaches is precisely \emph{because} diffusion models memorize more.

\subsection{Threat Model}

Our threat model considers an adversary $\mathcal{A}$ that interacts with a diffusion model $\gen$ (backed by a neural network $f_\theta$) to extract images from the model's training set $D$.

\tightparagraph{Image-generation systems.}
Unconditional diffusion models are trained on a dataset $D=\{x_1, x_2, \dots, x_n\}$. When queried, the system outputs a generated image $\xgen \gets \gen(r)$ using a fresh random noise $r$ as input.
Conditional models are trained on annotated images (e.g., labeled or captioned) $D=\{(x_1, c_1), \dots, (x_n, c_n)\}$ and when queried with a \emph{prompt} $p$, the system outputs  $\xgen \gets \gen(p; r)$ using the prompt $p$ and noise $r$.

\tightparagraph{Adversary capabilities.}
We consider two adversaries:
\begin{itemize}
    \item A \emph{black-box} adversary can query $\gen$ to generate images. If $\gen$ is a conditional generator, the adversary can provide arbitrary prompts $p$. The adversary cannot control the system's internal randomness $r$.
    \item A \emph{white-box} adversary gets full access to the system $\gen$ and its internal diffusion model $f_\theta$. They can control the model's randomness and can thus use the model to denoise arbitrary input images.
\end{itemize}

\noindent In both cases, we assume that an adversary who attacks a conditional image generator knows the captions for some images in the training set---thus allowing us to study the \emph{worst-case} privacy risk in diffusion models.

\tightparagraph{Adversary goals.}
We consider three broad types of adversarial goals, from strongest to weakest attacks:

\begin{enumerate}[itemsep=2pt]
    \item \emph{Data extraction}: The adversary aims to recover an image from the training set $x \in D$. The attack is successful if the adversary extracts an image $\hat{x}$ that is almost identical (see \Cref{ssec:memorization}) to \emph{some} $x \in D$. 
    
    \item \emph{Data reconstruction}: The adversary has partial knowledge of a training image $x \in D$ (e.g., a subset of the image) and aims to recover the full image. This is an image-analog of an \emph{attribute inference attack}~\cite{yeom2018privacy}, which aims to recover unknown features from partial knowledge of an input.
    
    \item \emph{Membership inference}: Given an image $x$, the adversary aims to infer whether $x$ is in the training set.
\end{enumerate}

\ifarxiv

\subsection{Ethics and Broader Impact}

Training data extraction attacks can present a threat to user privacy.
We take numerous steps to mitigate any possible harms from our paper.
First, we study models that are trained on publicly-available images (e.g., LAION and CIFAR-10) and therefore do not expose any data that was not already available online.

Nevertheless, data that is available online may not have been intended to be available online.
LAION, for example, contains unintentionally released medical images of several patients~\cite{laion_medical}.
We also therefore ensure that all images shown in our paper are of public figures (e.g., politicians, musicians, actors, or authors) who knowingly chose to place their images online.
As a result, inserting these images in our paper is unlikely to cause any unintended privacy violation.
For example, Figure~\ref{fig:teaser} comes from Ann Graham Lotz's Wikipedia profile picture and is licensed under Creative Commons, which allows us to ``redistribute the material in any medium'' and ``remix, transform, and build upon the material for any purpose, even commercially''.

Third, we shared an advance copy of this paper with the authors of each of the large-scale diffusion models that we study. This gave the authors and their corresponding organizations the ability to consider possible safeguards and software changes ahead of time.

In total, we believe that publishing our paper and publicly disclosing these privacy vulnerabilities is both ethical and responsible.
Indeed, at the moment, no one appears to be immediately harmed by the (lack of) privacy of diffusion models; our goal with this work is thus to make sure to preempt these harms and encourage responsible training of diffusion models in the future.


\fi

\section{Extracting Training Data from State-of-the-art Diffusion Models}
\label{sec:extraction}

We begin our paper by extracting training images from large, pre-trained, high-resolution diffusion models.
%

\subsection{Defining Image Memorization}
\label{ssec:memorization}

Most existing literature on training data extraction focuses on text language models, where a sequence is said to be ``extracted'' and ``memorized'' if an adversary can prompt the model to recover a \emph{verbatim} sequence from the training set~\cite{carlini2021extracting,kandpal2022deduplicating}.
Because we work with high-resolution images,
verbatim definitions of memorization are not suitable. 
%
%
Instead, we define a notion of approximate memorization based on image similarity metrics.

\newtheorem{definition}{Definition}

\begin{definition}[$(\ell,\errorthr)$-Diffusion Extraction] \label{definition:diffusion_extraction}
\emph{[adapted from~\cite{carlini2021extracting}]}.
We say that an example $x$ is \emph{extractable} from
a diffusion model $f_{\theta}$ if there exists an efficient algorithm $\mathcal{A}$
(that does not receive $x$ as input) such that $\hat{x} = \mathcal{A}(f_{\theta})$
has the property that $\ell(x, \hat{x}) \le \errorthr$.
\end{definition}


\noindent Here, $\ell$ is a distance function and $\errorthr$ is a threshold that determines whether we count two images as being identical.
In this paper, unless otherwise noted we follow 
Balle \emph{et al.}~\cite{balle2022reconstructing}
and use the Euclidean 2-norm distance
 $\ell_2(a, b) = \sqrt{\sum_{i} (a_i - b_i)^2 / d}$
where $d$ is the dimension of the inputs to normalize $\ell \in [0,1]$.
Given this definition of extractability, we can now define \emph{memorization}.

\begin{definition}[$(k,\ell,\errorthr)$-Eidetic Memorization] \emph{[adapted from~\cite{carlini2021extracting}]}.
We say that an example $x$ is $(k,\ell,\errorthr)$-Eidetic memorized \footnote{This paper covers a very restricted definition of ``memorization'': whether diffusion models can be induced to generate near-copies of some training examples when prompted with appropriate instructions. We will describe an approach that can generate images that are close approximations of some training images (especially images that are frequently represented in the training dataset through duplication or other means). There is active discussion within the technical and legal communities about whether the presence of this type of ``memorization'' suggests that generative neural networks ``contain'' their training data.} by a diffusion model
if $x$ is extractable from the diffusion model, and there are at most $k$
training examples $\hat x \in X$ where $\ell(x, \hat x) \le \errorthr.$
\label{definition:eidetic-mem}
\end{definition}

\noindent
Again, $\ell$ is a distance function and $\errorthr$ is its corresponding threshold.
The constant $k$ quantifies the number of near-duplicates of $x$ in the dataset. If $k$ is a small fraction of the data, then memorization is likely problematic. When $k$ is a larger fraction of data, memorization might be expected---but it could still be problematic, e.g., if the duplicated data is copyrighted.

\ifarxiv
\tightparagraph{Restrictions of our definition.} Our definition of extraction is intentionally conservative as compared to what privacy concerns one might ultimately have.
For example, if we prompt Stable Diffusion to generate ``A Photograph of Barack Obama,'' it produces an entirely recognizable photograph of Barack Obama but
not an \emph{near-identical reconstruction} of any particular training image.
Figure~\ref{fig:obama} compares the generated image (left) to the 4 nearest training images
under the Euclidean 2-norm (right).
Under our memorization definition, this image would not count as memorized. Nevertheless, the model's ability to generate (new) recognizable pictures of certain individuals could still cause privacy harms.
%

%

%

\begin{figure}
    \centering
    \includegraphics[scale=.2]{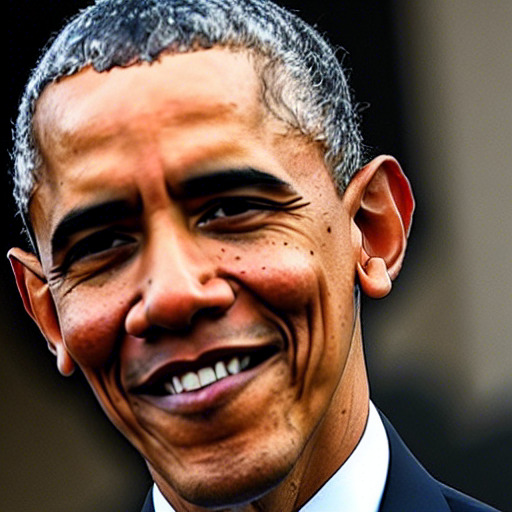}
    \includegraphics[scale=.2]{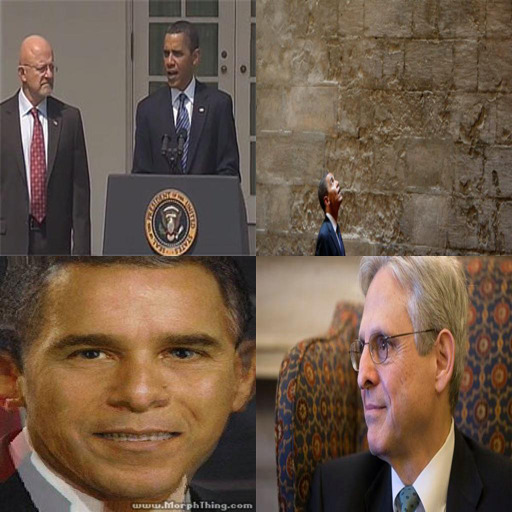}
    \caption{We do not count the generated
    image of Obama (at left) as memorized because it has
    a high $\ell_2$ distance to every training image.
    The four nearest training images are shown at right,
    each has a distance above $0.3$.
    }
    \label{fig:obama}
\end{figure}

\fi

\begin{figure*}
    \begin{tabular}{m{1.5cm}l}
     \begin{tabular}{r}\,\,\,\,Original:\end{tabular} \vspace{1.3em}&  \includegraphics[width=1.15cm]{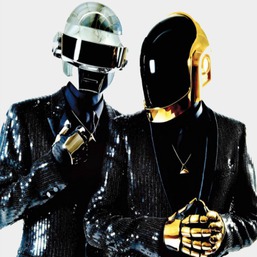}\includegraphics[width=1.15cm]{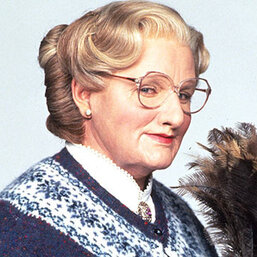}\includegraphics[width=1.15cm]{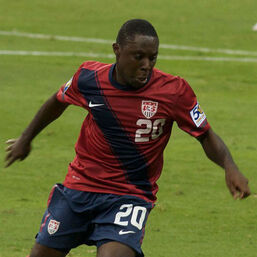}\includegraphics[width=1.15cm]{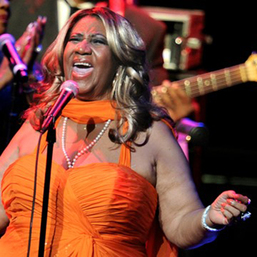}\includegraphics[width=1.15cm]{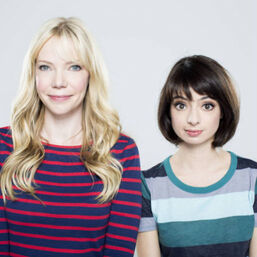}\includegraphics[width=1.15cm]{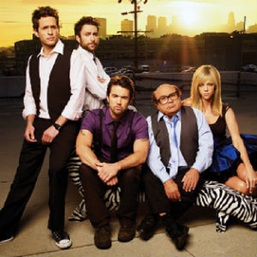}\includegraphics[width=1.15cm]{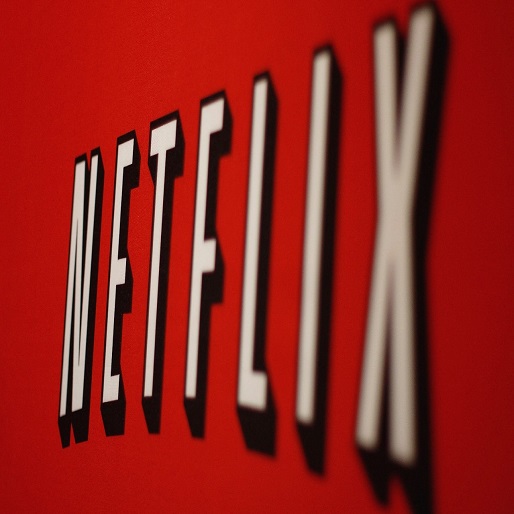}\includegraphics[width=1.15cm]{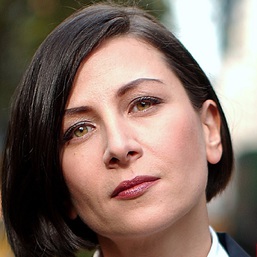}\includegraphics[width=1.15cm]{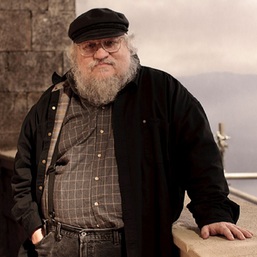}\includegraphics[width=1.15cm]{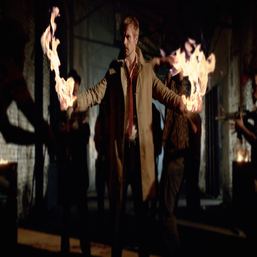}\includegraphics[width=1.15cm]{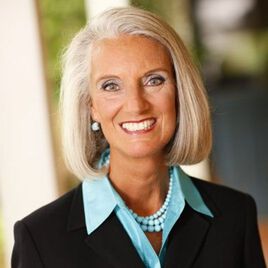}\includegraphics[width=1.15cm]{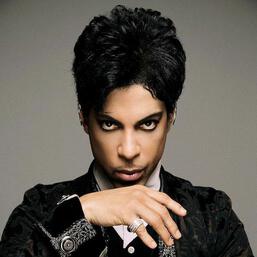}\\
     \begin{tabular}{r}Generated:\end{tabular} \vspace{1.3em}& \includegraphics[width=1.15cm]{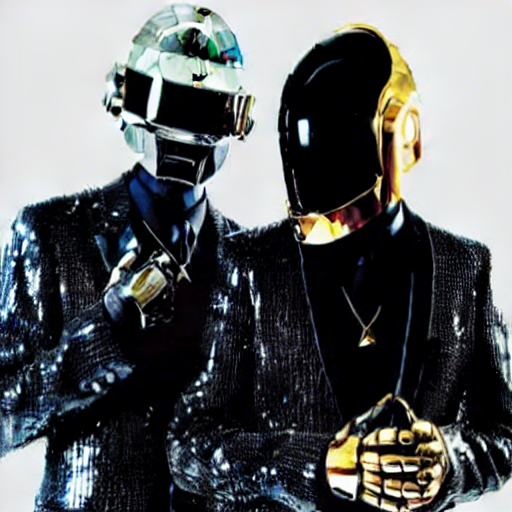}\includegraphics[width=1.15cm]{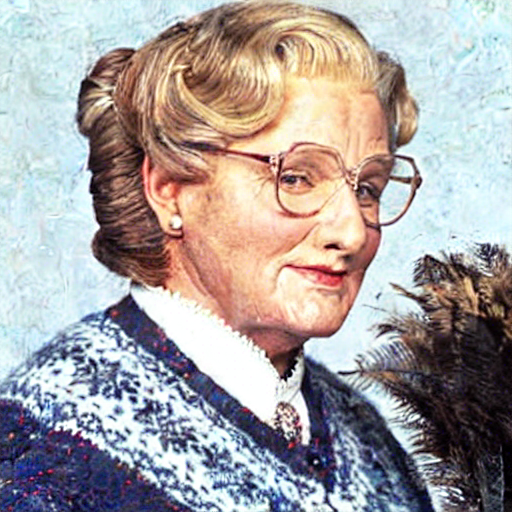}\includegraphics[width=1.15cm]{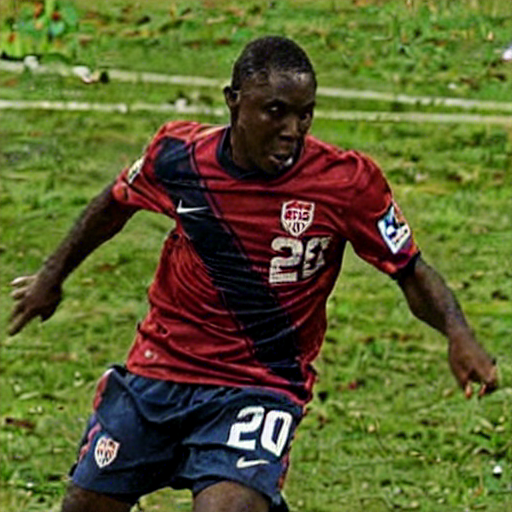}\includegraphics[width=1.15cm]{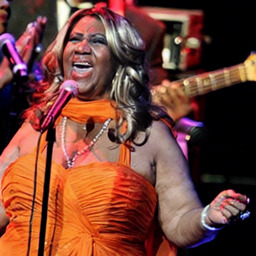}\includegraphics[width=1.15cm]{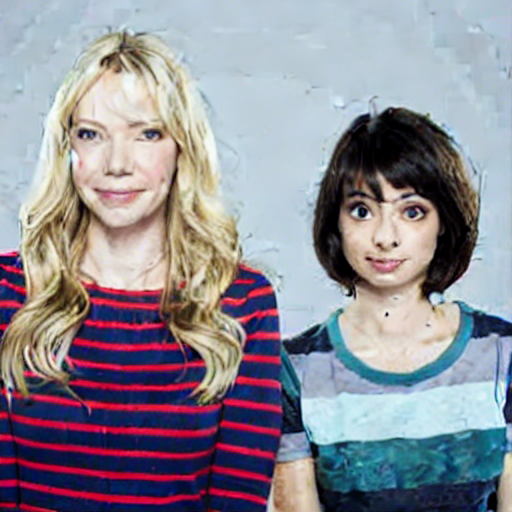}\includegraphics[width=1.15cm]{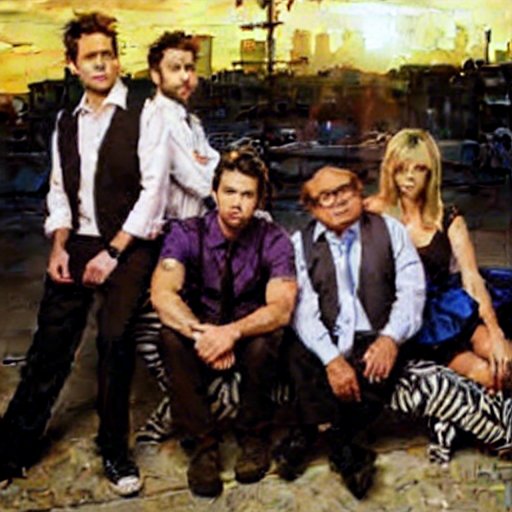}\includegraphics[width=1.15cm]{figures/stable_diffusion_v1_demo/198848_original.jpeg}\includegraphics[width=1.15cm]{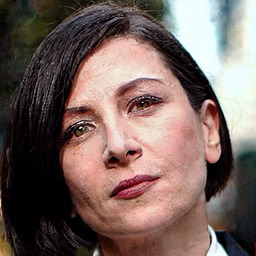}\includegraphics[width=1.15cm]{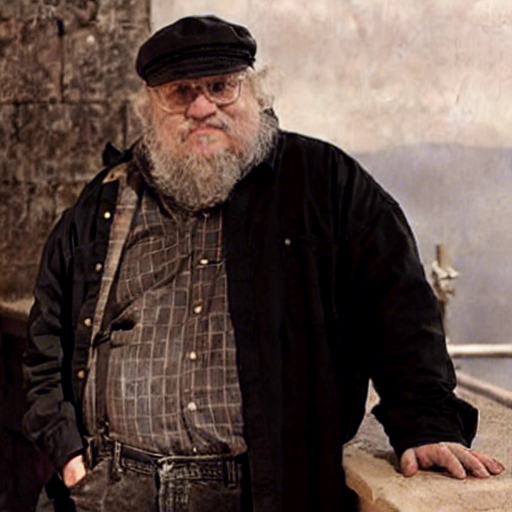}\includegraphics[width=1.15cm]{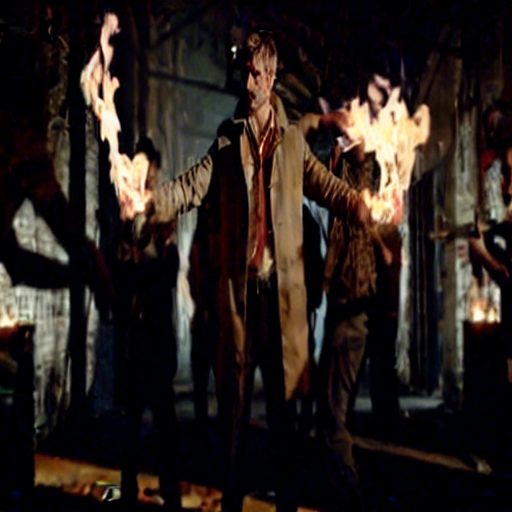}\includegraphics[width=1.15cm]{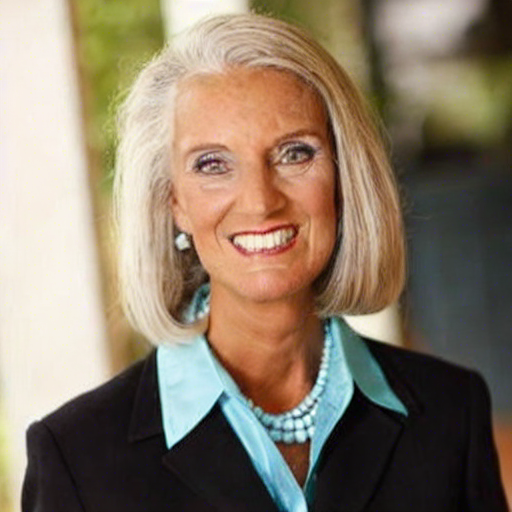}\includegraphics[width=1.15cm]{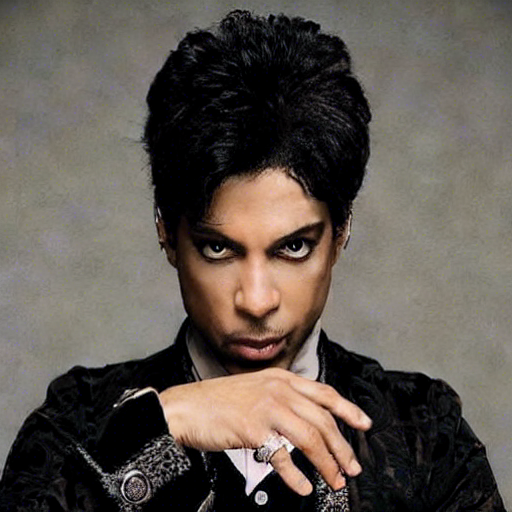}
\end{tabular}

    \vspace{-0.65cm}
    \caption{Examples of the images that we extract from Stable Diffusion v1.4 using random sampling and our membership inference procedure. 
    The top row shows the original images and the bottom row shows our extracted images.}
    \label{fig:sd_14_extractions_sample}
\end{figure*}

\subsection{Extracting Data from Stable Diffusion}\label{ssec:stablediffusion}

We now extract training data from Stable Diffusion:
the largest and most popular open-source diffusion model ~\cite{rombach2022high}.
This model is an 890 million parameter text-conditioned diffusion model trained on 160 million images. We generate from the model using the default PLMS sampling scheme at a resolution of $512\times512$ pixels.
As the model is trained on publicly-available images, we can easily verify our attack's success and also mitigate potential harms from exposing the extracted data.
We begin with a black-box attack.

\tightparagraph{Identifying duplicates in the training data.} 
To reduce the computational load of our attack,
as is done in \cite{somepalli2022diffusion},
we bias our search towards duplicated training examples
because these are orders of
magnitude more likely to be memorized than non-duplicated examples~\cite{lee2021deduplicating,kandpal2022deduplicating}.
%
%
%

If we search for images that are bit-for-bit identically duplicated in the
training dataset, we would significantly undercount the true rate of duplication.
Instead, we account for near-duplication.
Ideally, we would search for any training examples
that are nearly duplicated with a pixel-level $\ell_2$ distance below some threshold. But this is computationally intractable, as it would
require an all-pairs comparison of 160 million images in
Stable Diffusion's training set, each of which is a $512 \times 512 \times 3$ dimensional vector.
Instead, we first \emph{embed} each image to a $512$ dimensional vector
using CLIP~\cite{radford2021learning},
and then perform the all-pairs comparison between images in this lower-dimensional
space (increasing efficiency by over $1500\times$).
We count two examples as near-duplicates if their CLIP embeddings have a high cosine similarity.
For each of these near-duplicated images, we use the corresponding captions as the input to our extraction attack.

\subsubsection{Extraction Methodology}
\label{sssec:extracting_duplicated_images}

%
Our extraction approach adapts the methodology from prior work 
\cite{carlini2021extracting} to images and consists of two steps:
\begin{enumerate}
    \item \emph{Generate many examples} using the diffusion model in the standard
    sampling manner and with the known prompts from the prior section.
    
    \item \emph{Perform membership inference}
    to separate the model's novel generations 
    from those generations which are memorized training examples.
\end{enumerate}

\tightparagraph{Generating many images.} The first step is trivial but computationally expensive:
we query the $\gen$ function in a black-box manner using the selected
prompts as input.
To reduce the computational overhead of our experiments, we use the
timestep-resampled generation implementation that is available in the Stable Diffusion codebase \cite{rombach2022high}.
This process generates images in a more aggressive fashion by removing larger amounts of noise at each time step
%
and results in slightly lower visual fidelity at a significant ($\sim 10\times$) performance increase.
We generate $500$ candidate images for each text prompt
to increase the likelihood that we find memorization.

\tightparagraph{Performing membership inference.} The second step requires flagging generations that appear to be memorized training images.
Since we assume a black-box threat model in this section, we do not have access to the loss and cannot exploit techniques from state-of-the-art membership inference attacks~\cite{carlini2021extracting}.
%
%
We instead design a new membership inference attack strategy based on the intuition that for diffusion models, with high probability $\gen(p; r_1) \ne \gen(p; r_2)$ for two different
random initial seeds $r_1, r_2$.
On the other hand, if $\gen(p; r_1) \approx_{d} \gen(p; r_2)$ under some distance
measure $d$, 
it is likely that these generated samples are memorized examples.

The 500 images that we generate for each prompt have different (but unknown) random seeds.
We can therefore construct a graph over the $500$ generations by connecting an edge
between generation $i$ and $j$ if $x_i \approx_{d} x_j$.
If the largest clique in this graph is at least size 10 (i.e., $\geq$ 10 of the 500 generations are near-identical), we predict that this clique is a memorized image.
Empirically, clique-finding is more effective than searching for \emph{pairs} of images $x_1 \approx_{d} x_2$ as it has fewer false positives.

To compute the distance measure $d$ among the images in the clique, we use a modified Euclidean $\ell_2$ distance. In particular, we found that many generations were often spuriously similar according to $\ell_2$ distance (e.g., they all had gray background). We therefore instead divide each image into 16 non-overlapping $128\times128$ tiles and measure the maximum of the $\ell_2$ distance between any pair of image tiles between the two images. 




\begin{figure}
    \centering
    \includegraphics[scale=.75]{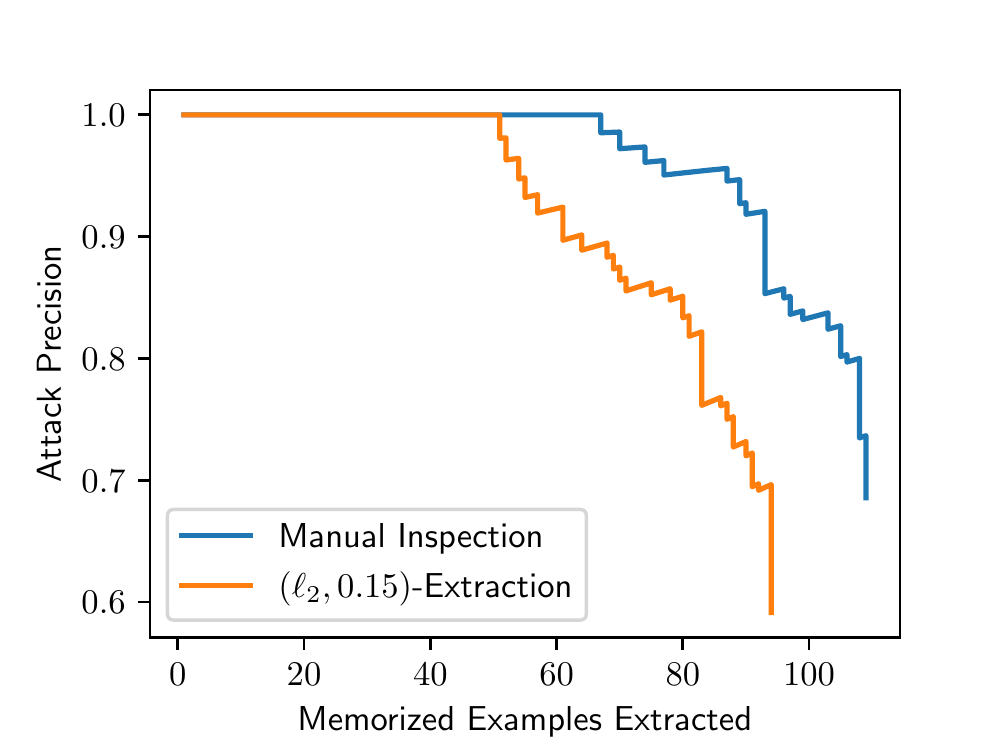}
    \vspace{-0.2cm}
    \caption{Our attack reliably separates novel generations from memorized training examples,
    under two definitions of memorization---either $(\ell_2, 0.15)$-extraction or manual human inspection of generated images.
    }
    \label{fig:prcurve}
\end{figure}

\subsubsection{Extraction Results}
In order to evaluate the effectiveness of our attack,
we select the 350,000 most-duplicated examples from the training dataset
and generate 500 candidate images for each of these prompts (totaling 175 million generated images).
We first sort all of these generated images 
by ordering them by the mean distance between images in the clique to identify generations that we predict are likely to be memorized training data.
We then take each of these generated images and annotate each as either ``extracted'' or ``not extracted'' by comparing it to the training images under Definition~\ref{definition:diffusion_extraction}.
%
We find 94 images are $(\ell_2, 0.15)$-extracted.
To ensure that these images not only match some arbitrary definition,
we also manually annotate the top-1000 generated images as either memorized or not memorized by visual analysis, and find that a further 13 (for a total of 109 images) are near-copies of training examples
even if they do not fit our 2-norm definition.
Figure \ref{fig:sd_14_extractions_sample} shows a subset of the extracted images
that are reproduced with near pixel-perfect accuracy;
all images have an $\ell_2$ difference under $0.05$.
(As a point of reference, re-encoding a PNG as a JPEG with quality level 50
results in an $\ell_2$ difference of $0.02$ on average.)

Given our ordered set of annotated images, we can also compute
a curve evaluating the number of extracted images
to the attack's false positive rate.
Our attack is exceptionally precise: out of 175 million generated images, 
we can identify $50$ memorized images with $0$ false positives,
and all our memorized images can be extracted with a precision above $50\%$.
Figure~\ref{fig:prcurve} contains the precision-recall curve for both memorization definitions.

%


%

\begin{figure}[t]
    \centering
    \vspace{-0.2cm}
    \includegraphics[scale=0.75]{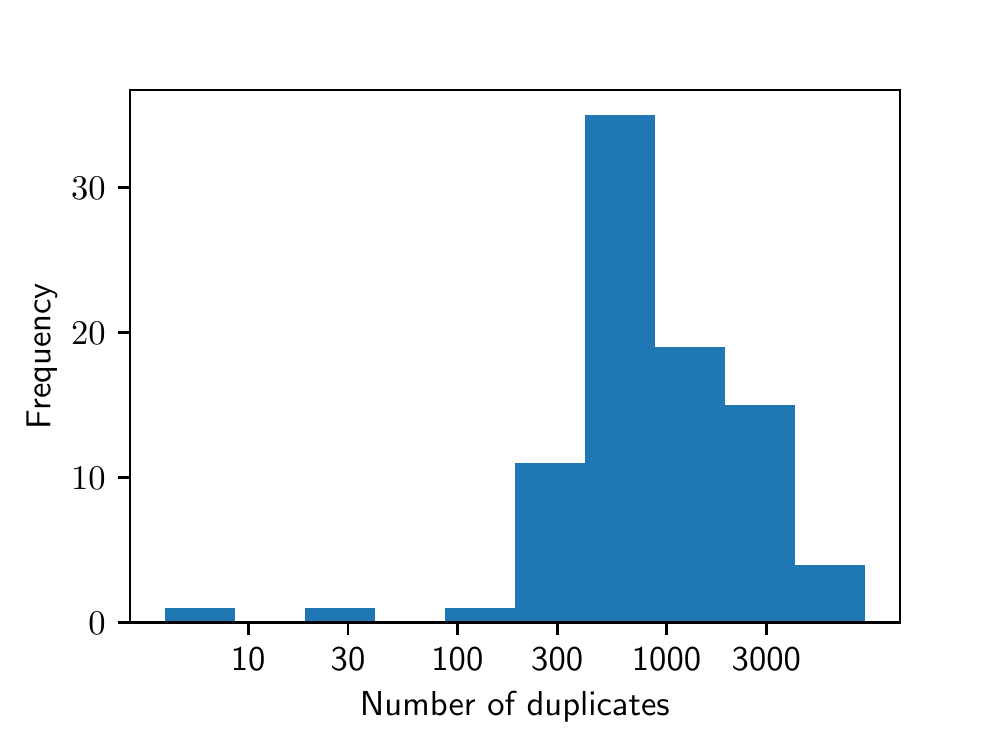}
    \caption{Our attack extracts images from Stable Diffusion
    most often when they have been duplicated at least $k=100$ times;
    although this should be taken as an upper bound because our methodology
    explicitly searches for memorization of duplicated images.}
    \label{fig:sd_14_rep_histogram}
\end{figure}

\paragraph{Measuring $(k,\ell,\errorthr)$-eidetic memorization.}
In Definition \ref{definition:eidetic-mem} we introduced an adaptation of Eidetic memorization \cite{carlini2021extracting}
tailored to the domain of generative image models.
As mentioned earlier, we compute similarity between pairs of images with a direct $\ell_2$ pixel-space similarity.
This analysis is computationally expensive\footnote{In practice it is even more challenging:
for non-square images, Stable Diffusion takes a random square crop,
and so to check if the generated image $x$ matches a non-square training image $y$
we must try all possible alignments between $x$ on top of the image $y$.} as it requires comparing each of our
memorized images against each of the $160$ million training examples.
We set $\errorthr=0.1$ as this threshold is sufficient to identify almost all small image corruptions (e.g., JPEG compression, small brightness/contrast adjustments) but has very few false positives.

Figure~\ref{fig:sd_14_rep_histogram} shows the results of this analysis.
While we identify little Eidetic memorization for $k<100$,
this is expected due to the fact we choose prompts of highly-duplicated images.
Note that at this level of duplication, the duplicated examples still make up
just \emph{one in a million} training examples. These results show that duplication is a major factor behind training data extraction.

\paragraph{Qualitative analysis.}
%
The majority of the images that we extract (58\%) are photographs with a recognizable person
as the primary subject; the remainder are mostly either
products for sale (17\%), logos/posters (14\%), or other art or graphics.
We caution that if a future diffusion model were trained on
sensitive (e.g., medical) data, then the kinds of data that we extract
would likely be drawn from this sensitive data distribution.

Despite the fact that these images are publicly accessible on the Internet,
not all of them are permissively licensed.
We find that a significant number of these images fall under
an explicit non-permissive copyright notice (35\%). Many other images (61\%) have no explicit copyright notice but may fall under a general copyright protection for the website that hosts them (e.g., images of products on a sales website).
Several of the images that we extracted are licensed CC BY-SA, which requires
``[to] give appropriate credit, provide a link to the license, and indicate if changes were made.''
Stable Diffusion thus memorizes numerous copyrighted and non-permissive-licensed images, which the model may reproduce without the accompanying license.




\subsection{Extracting Data from Imagen}

While Stable Diffusion is the best publicly-available diffusion model,
there are non-public models that achieve stronger performance
 using larger models and datasets~\cite{ramesh2022hierarchical,saharia2022photorealistic}.
Prior work has found that larger models are more likely to memorize training
data~\cite{carlini2021extracting,carlini2022quantifying} and we thus study Imagen~\cite{saharia2022photorealistic}, 
a 2 billion parameter text-to-image diffusion model.
While individual details differ between Imagen's and Stable Diffusion's
implementation and training scheme, these details are independent of our extraction results.


We follow the same procedure as earlier but focus on the top-1000 most duplicated prompts
for computational reasons.
We then generate 500 images for each of these prompts,
and compute the $\ell_2$ similarity between each generated image and the
corresponding training image.
By repeating the same membership inference steps as above---searching
for cliques under patched $\ell_2$ distance--we identify 23 of these
$1{,}000$ images as memorized training examples.\footnote{Unfortunately, because the Imagen training dataset is not public, 
we are unable to provide visual examples of successful reconstructions.}
This is significantly higher than the rate of memorization in Stable Diffusion,
and clearly demonstrates that memorization across diffusion models is
highly dependent on training settings such as the model size, training time, and dataset size.


\subsection{Extracting Outlier Examples}
\label{sec:ood}

The attacks presented above succeed, 
but only at extracting images that are highly duplicated.
This ``high $k$'' memorization may be problematic,
but as we mentioned previously, the most compelling
practical attack would be to demonstrate memorization
in the ``low $k$'' regime.
%

We now set out to achieve this goal.
In order to find non-duplicated examples likely to be memorized,
we take advantage of the fact that while on \emph{average} models often respect
the privacy of the majority of the dataset, there often exists a small set of ``outlier''
examples whose privacy is more significantly exposed~\cite{feldman2020does}.
And so instead of searching for memorization across all images,
we are more likely to succeed if we focus our effort on these outlier
examples.

But how should we find which images are potentially outliers?
Prior work was able to train hundreds of models on subsets of the training dataset and then use
an influence-function-style approach to identify examples that
have a significant impact on the final model weights~\cite{feldman2020neural}.
Unfortunately, given the cost of training even a single large
diffusion model is in the millions-of-dollars,
this approach will not be feasible here. 

Therefore we take a simpler approach.
We first compute the CLIP embedding of each training example,
and then compute the ``outlierness'' of each example as the
average distance (in CLIP embedding space) to its $1{,}000$ nearest neighbors in the training dataset.

\paragraph{Results.}
Surprisingly, we find that attacking out-of-distribution images is much
more effective for Imagen than it is for Stable Diffusion. 
On Imagen, we attempted extraction of the 500 images with the highest out-of-distribution score.
Imagen memorized and regurgitated 3 of these images (which were \emph{unique} in the training dataset).
In contrast, we failed to identify \emph{any} memorization when applying the same methodology to Stable Diffusion---even after attempting to extract the $10{,}000$ most-outlier samples.
Thus, Imagen appears less private than Stable Diffusion both on duplicated and non-duplicated images.
We believe this is due to the fact that Imagen uses a model with a much higher capacity compared to Stable diffusion, which allows for more memorization~\cite{carlini2022quantifying}. Moreover, Imagen is trained for more iterations and on a smaller dataset, which can also result in higher memorization.

\section{Investigating Memorization}
\label{sec:cifar10}


The above experiments are visually striking and clearly indicate that memorization is
pervasive in large diffusion models---and that data extraction is feasible.
But these experiments do not explain
\emph{why} and \emph{how} these models memorize training data.
In this section we train smaller diffusion models
and perform controlled experiments in order to
more clearly understand memorization.

\paragraph{Experimental setup.}
For the remainder of this section, we focus on diffusion models trained on CIFAR-10.
We use state-of-the-art training code \footnote{We either directly use OpenAI's Improved Diffusion repository (\texttt{https://github.com/openai/improved-diffusion}) in 
\Cref{ssec:cifar10_extract}, or our own re-implementation in all following sections. Models trained with our re-implementation achieve almost identical FID to the open-sourced models. We use half the dataset as is standard in privacy analyses~\cite{carlini2022membership}.} 
to train 16 diffusion models,
each on a randomly-partitioned half
of the CIFAR-10 training dataset.
We run three types of privacy attacks: membership inference attacks, attribute inference attacks, and data reconstruction attacks.
For the membership inference attacks, we train class-conditional models that reach an FID below 3.5 (see \Cref{fig:mia_fid_tpr_fpr}), placing them in
the top-30 generative models on CIFAR-10 \cite{paperswithcode}.
For reconstruction attacks (\Cref{ssec:cifar10_extract}) and attribute inference attacks with inpainting (\Cref{ssec: cifar10_inpaint}), we train unconditional models with an FID below 4.

\subsection{Untargeted Extraction}
\label{ssec:cifar10_extract}

Before devling deeper into understanding memorization,
we begin by validating that memorization does still occur in our smaller models.
Because these models are not text conditioned, we focus on \emph{untargeted} extraction.
Specifically, given our $16$ diffusion models trained on CIFAR-10, we unconditionally generate
$2^{16}$ images from each model for a total of $2^{20}$ candidate images.
Because we will later develop high-precision membership inference attacks,
in this section we directly search for memorized training examples among all our million
generated examples.
Thus this is not an attack \emph{per se}, but rather verifying the
capability of these models to memorize.
%

\begin{figure}
    \centering
    \includegraphics[scale=.55]{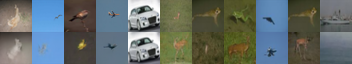}
    \caption{Direct 2-norm measurement fails to identify memorized CIFAR-10 examples.
    Each of the above images have a $\ell_2$ distance of less than $0.05$,
    yet only one (the car) is actually a memorized training example.
    }
    \label{fig:badl2}
\end{figure}

\tightparagraph{Identifying matches.}
In the prior section, we performed targeted attacks and could therefore check for successful memorization by simply computing the $\ell_2$ distance between the target image and the generated image.
Here, as we perform an all-pairs comparison, we find that using an uncalibrated $\ell_2$ threshold
fails to accurately identify memorized training examples.
For example, if we set a highly-restrictive threshold of $0.05$, then
nearly all ``extracted'' images are of entirely blue skies or green landscapes (see Figure~\ref{fig:badl2}).
We explored several other metrics (including perceptual distances like SSIM or CLIP embedding distance) but found that none could reliably identify memorized training images for CIFAR-10.

\begin{figure}
    \centering
    \vspace{-.5cm}
    \begin{overpic}[abs,unit=1mm,scale=.6]{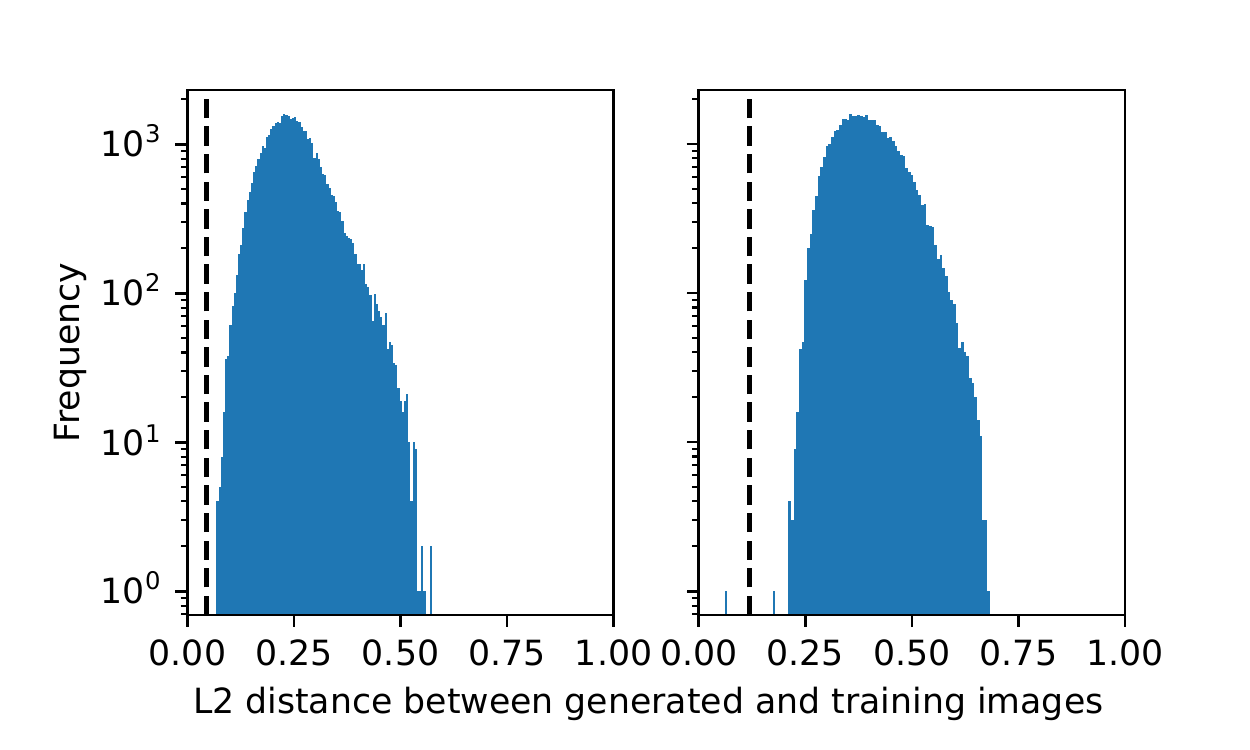}
       \put(29,23){\includegraphics[scale=.75]{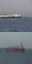}}
       \put(60,23){\includegraphics[scale=.75]{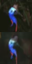}}
    \end{overpic}
    \caption{Per-image $\ell_2$ thresholds are necessary to separate
    memorized images from novel generations on a CIFAR-10 model.  
    Each plot shows the distribution of $\ell_2$ distances from a generated image to all training images (along with the image and the nearest training image).
    \textbf{Left} shows a typical distribution for a non-memorized image.
    \textbf{Right} shows a memorized image distribution;
    while the most similar training image has high absolute $\ell_2$ distance, 
    it is \emph{abnormally} low for this distribution.
    The dashed black line shows our adaptive $\ell_2$ threshold.}
    \label{fig:lossdistribution}
\end{figure}

%
We instead define an image as extracted if the $\ell_2$ distance to its nearest neighbor in the training set is \emph{abnormally low} compared to all other training images.
Figure~\ref{fig:lossdistribution} illustrates this by computing the $\ell_2$ distance between two different generated images and
every image in the CIFAR-10 training dataset.
The left figure shows a failed extraction attempt;
despite the fact that the nearest training image has an $\ell_2$ distance of just $0.06$,
this distance is on par with the distance to many other training images (i.e., all images that contain a blue sky).
In contrast, the right plot shows a successful extraction attack.
Here, even though the $\ell_2$ distance to the nearest training image is higher than
for the prior failed attack ($0.07$), this value is \emph{unusually small}
compared to other training images which almost all are at a distance above $0.2$.

We thus slightly modify our attack to use the distance
\[\ell(\hat{x}, x; S_{\hat{x}}) = {\ell_2(\hat{x}, x) 
\over 
{\alpha \cdot \mathbb{E}_{y \in S_{\hat{x}}} [ \ell_2(\hat{x}, y) ]}}.\]
where $S_{\hat{x}}$ is the set containing the $n$ closest 
elements from the training dataset to the example $\hat{x}$.
This distance is small if the extracted image $x$ is much closer to the training image
$\hat{x}$ compared to the $n$ closest neighbors of $\hat{x}$ in the training set.
We run our attack with $\alpha=0.5$ and $n=50$. Our attack was not
sensitive to these choices.

\begin{figure*}
    \centering
   \includegraphics[scale=.8]{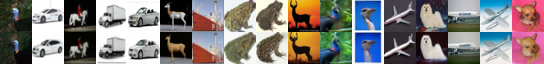}
    \caption{Selected training examples that we extract from a diffusion model trained on CIFAR-10 by sampling from the model 1 million times.
    \textbf{Top} row: generated output from a diffusion model.
    \textbf{Bottom} row: nearest ($\ell_2)$ example from the training dataset.
    Figure~\ref{fig:cifar_all_extracted} in the Appendix contains all $1{,}280$ unique extracted images.}
    \label{fig:extracted_selected}
\end{figure*}

\tightparagraph{Results.}
Using the above methodology we identify $1{,}280$ unique
extracted images from the CIFAR-10 dataset ($2.5\%$ of the entire dataset).\footnote{
Some CIFAR-10 training images are generated multiple times.
In these cases, we only count the first generation as a successful attack.
Further, because the CIFAR-10 training dataset contains many duplicate images,
we do not count two generations of two different (but duplicated) images in the training dataset.}
In Figure~\ref{fig:extracted_selected} we show a selection of training examples that we extract and full results are shown in Figure~\ref{fig:cifar_all_extracted} in the Appendix.
%

\subsection{Membership Inference Attacks}
\label{ssec:cifar_mi}
We now evaluate membership inference with more traditional attack techniques
that use white-box access, as opposed to
Section \ref{sssec:extracting_duplicated_images} that assumed black-box access.
We will show that \emph{all} examples have significant privacy leakage under membership inference attacks, compared to the small fraction that are sensitive to data extraction.
We consider two membership inference attacks on our class-conditional CIFAR-10-trained diffusion models.\footnote{\Cref{ssec:mia_cond_vs_uncond} replicates these results for unconditional models.}

\tightparagraph{The loss threshold attack.}
Yeom \textit{et al.} \cite{yeom2018privacy} introduce the simplest membership inference attack:
because models are trained to minimize their loss on the training set, we should expect that training examples
have lower loss than non-training examples.
%
The loss threshold attack thus computes the loss $l = \mathcal{L}(x; f)$ and reports
``member'' if $l < \tau$ for some chosen threshold $\tau$ and otherwise ``non-member'.
The value of $\tau$ can be selected to maximize a desired metric (e.g., true positive rate at some
fixed false positive rate or the overall attack accuracy).

\tightparagraph{The Likelihood Ratio Attack (LiRA).}
Carlini \textit{et al.} \cite{carlini2022membership} introduce the state-of-the-art 
approach to performing membership inference attacks.
LiRA first trains a collection of \emph{shadow models},
each model on random subsets of the training dataset.
LiRA then computes the loss $\mathcal{L}(x; f_i)$ for
the example $x$ under each of these shadow models $f_i$.
These losses are split into two sets: the losses $\texttt{IN}=\{l^{\text{in}_i}\}$ for the example $x$
under the shadow models $\{f_i\}$ that \emph{did} see the example $x$ during training,
and the losses $\texttt{OUT}=\{l^{\text{out}_i}\}$ for the example $x$ under the shadow models $\{f_j\}$ that \emph{did not}
see the example $x$ during training.
LiRA finishes the initialization process by fitting Gaussians $N_{IN}$ to the \texttt{IN} set and $N_{OUT}$ to \texttt{OUT}
set of losses.
Finally, to predict membership inference for a new model $f^*$, we compute 
$l^* = \mathcal{L}(x, f^*)$ and then measure whether $Pr[l^* | N_{IN}] > Pr[l^* | N_{OUT}]$.

\tightparagraph{Choosing a loss function.}
Both membership inference attacks use a loss function $\mathcal{L}$.
In the case of classification models,
Carlini \emph{et al.}~\cite{carlini2022membership} find that choosing a loss function is one of the most
important components of the attack. 
We find that this effect is even more pronounced for diffusion models.
In particular, unlike classifiers that have a single loss function (e.g., cross entropy)
used to train the model,
diffusion models are trained to minimize the reconstruction loss
when a random quantity of Gaussian noise $\epsilon$ has been added to an image.
This means that ``the loss'' of an image is not well defined---instead,
we can only ask for the loss $\mathcal{L}(x, t, \epsilon)$
of an image $x$ for a certain timestep $t$ with a corresponding amount of noise $\epsilon$ (cf.\ \Cref{eqn:diffusion_loss}).

We must thus compute the optimal timestep $t$ at which we should measure the loss. To do so,
we train 16 shadow models each on a random
50\% of the CIFAR-10 training dataset.
We then compute the loss for every model, for every example in the training dataset, and every timestep $ t\in[1, T]$ ($T = 1{,}000$ in the models we use).

\begin{figure}[t]
\centering
    \includegraphics[width=0.90\linewidth]{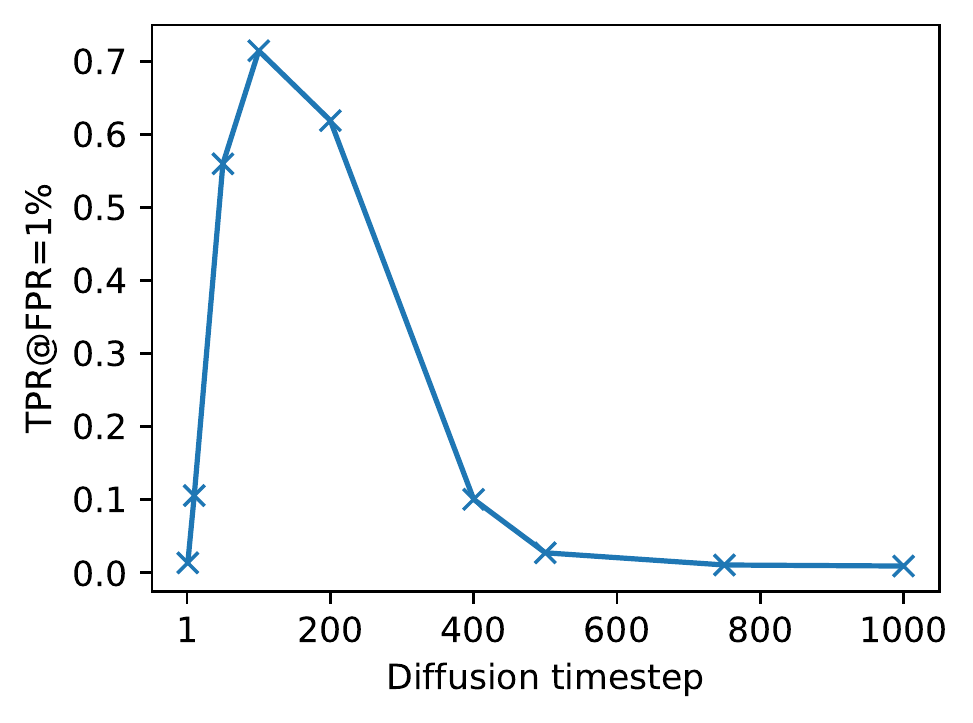}
\vspace{-0.2cm}
\caption{We run membership inference using LiRA and compute the diffusion model loss at different noise timesteps on CIFAR-10. Evaluating $\mathcal{L}(\cdot, t, \cdot)$ at $t\in[50, 300]$ produces the best results.}
\label{fig:mia_vs_diffusion_time_cifar10}
\end{figure}

\Cref{fig:mia_vs_diffusion_time_cifar10} plots the timestep used to compute the loss against the attack success rate, measured as the true positive rate (TPR), i.e., the number of examples which truly are members over the total number of members, at a fixed false positive rate (FPR) of 1\%, i.e., the fraction of examples which are incorrectly identified as members.
Evaluating $\mathcal{L}$ at $t\in[50, 300]$ leads to the most successful attacks.
We conjecture that this a ``Goldilock's zone'' for membership inference: if $t$ is too small, and so the noisy image is similar to the original, then predicting the added noise is easy regardless if the input was in the training set;
if $t$ is too large, and so the noisy image is similar to Gaussian noise, then the task is too difficult.
Our remaining experiments will evaluate $\mathcal{L}(\cdot,t,\cdot)$ at $t=100$, where we observed a TPR of 71\% at an FPR of 1\%.

\subsubsection{Baseline Attack Results}
\label{sec:baseline}

\begin{figure}[t]
  \centering
\centering
    \includegraphics[width=0.9\linewidth]{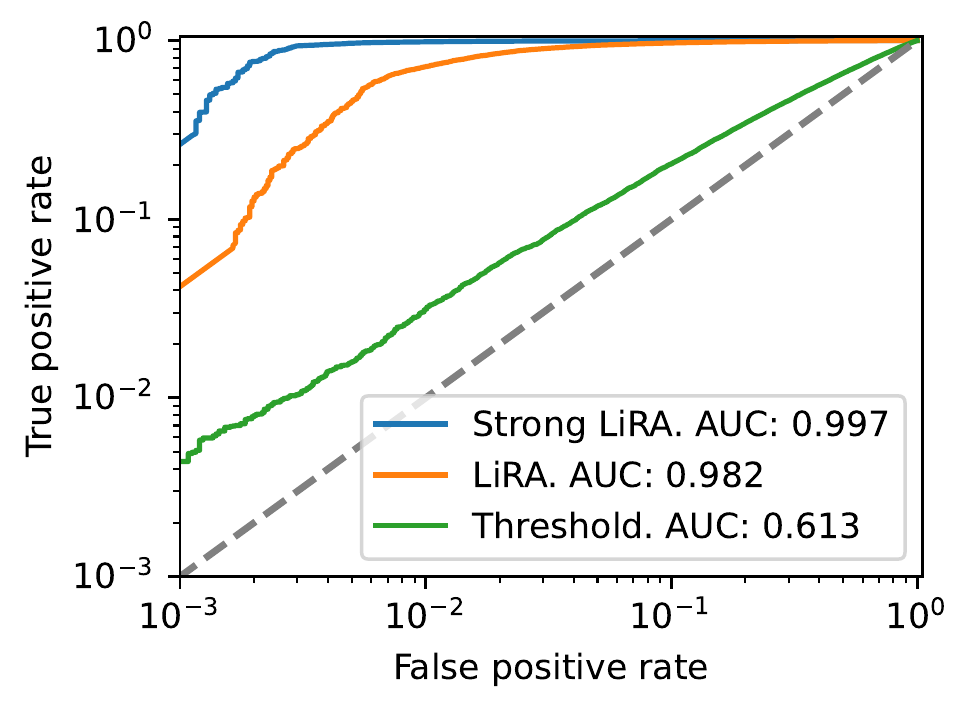}
        \vspace{-0.2cm}
        \caption{Membership inference ROC curve for a diffusion model trained on CIFAR-10 using the loss threshold attack, baseline LiRA, and ``Strong LiRA'' with repeated queries and augmentation (\S\ref{sec:augment}).}
    \label{fig:roc_curve}
\label{fig:compare_lira_mia}
\end{figure}

We now evaluate membership inference using our specified loss function.
We follow recent advice~\cite{carlini2022membership} and evaluate the efficacy of membership inference attacks
by comparing their true positive rate to the false positive rate 
on a log-log scale.
In~\Cref{fig:compare_lira_mia}, 
we plot the membership inference ROC curve for the loss threshold attack and LiRA.
An out-of-the-box implementation of LiRA achieves a true positive rate of over $70\%$ at a false positive rate of just $1\%$.
As a point of reference, state-of-the-art \emph{classifiers} are much more
private, e.g., with a $<20\%$ TPR at $1\%$ FPR~\cite{carlini2022membership}. This shows that diffusion models are significantly less private than classifiers trained on the same data.
(In part this may be because diffusion models are often trained far longer than classifiers.)

\ifarxiv
\tightparagraph{Qualitative analysis.}
In Figure~\ref{fig:inliers}, we visualize the least- and most-private images as
determined by their easiness to detect via LiRA.
%
%
We find that the easiest-to-attack examples are all extremely out-of-distribution
visually from the CIFAR-10 dataset.
These images are even more visually
out-of-distribution compared to the outliers identified by Feldman \emph{et al.}~\cite{feldman2020does} 
who produce a similar set of images but for image \emph{classifiers}.
In contrast, the images that are hardest to attack are \emph{all}
duplicated images.
It is challenging to detect the presence or absence of each of these images
in the training dataset because there is another \emph{identical} image in the training
dataset that may have been present or absent---therefore making the membership inference
question ill-defined.
\fi

\subsubsection{Augmentations Improve Attacks}
\label{sec:augment}

Membership inference attacks can also be improved by reducing the variance
in the loss signal~\cite{carlini2022membership, ye2021enhanced}.
We study two ways to achieve this for diffusion models.
First, because our loss function has randomness 
(recall that to compute the reconstruction loss we measure the quantity
$\mathcal{L}(x, t, \epsilon)$ for a random noise sample $\epsilon \sim \mathcal{N}(0,I)$),
we can compute a better estimate of the true loss by averaging over different noise samples:
$\mathcal{L}(x, t) = \mathbb{E}_{\epsilon \sim \mathcal{N}(0, I)} [\mathcal{L}(x, t, \epsilon)]$.

By varying the number of point samples taken to estimate this expectation
we can potentially increase the attack success rate.
And second, because our diffusion models train on \emph{augmented} versions
of training images (e.g., by flipping images horizontally), 
it makes sense to compute the loss averaged over all possible
augmentations.
Prior work has found that both of these attack strategies are effective at
increasing the efficacy of membership inference attacks for classifiers~\cite{carlini2022membership, jayaraman2020revisiting},
and we find they are effective here as well.

\tightparagraph{Improved attack results.} Figure~\ref{fig:compare_lira_mia} shows the effect of combining both these strategies. 
Together they are remarkably successful, and
at a false positive rate of $0.1\%$ they increase the true positive rate by over
a factor of six from $7\%$ to $44\%$.
\Cref{fig:improved_mia_with_repeated_noise} in the Appendix breaks down the impact of each component:
in~\Cref{fig:mia_repeated_noise} we increase
the number of Monte Carlo samples from 1 (the base LiRA attack) to 20, and in~\Cref{fig:mia_repeated_noise_and_flip} we
augment samples with a horizontal flip.
%
%

%

\subsubsection{Memorization Versus Utility} 
\label{ssec:mia_vs_fid}

We train our diffusion models to reach state-of-the-art levels of performance.
Prior work on language models has found that better models are often
\emph{easier} to attack than less accurate models---intuitively, because they extract more information from the same training dataset \cite{carlini2022quantifying}.
Here we perform a similar experiment.

\begin{figure}[t]
\centering
    \includegraphics[width=1.\linewidth]{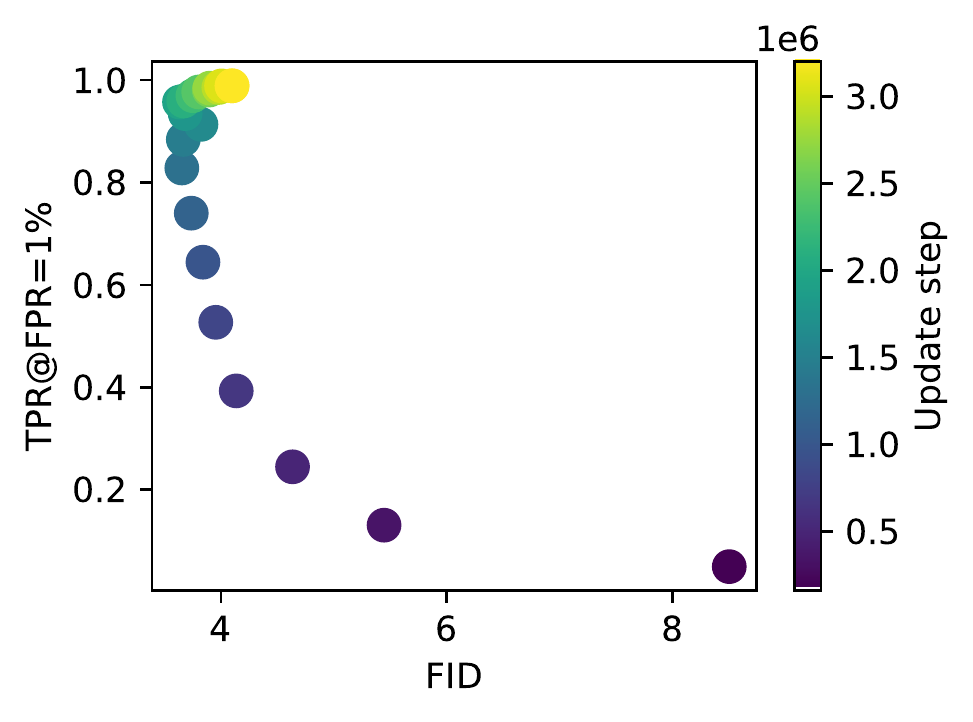}
\caption{Better diffusion models are more vulnerable to membership inference attacks; evaluating with TPR at an FPR of 1\%.
As the FID decreases (corresponding to a quality increase) the membership inference
attack success rate grows from $7\%$ to nearly $100\%$.}
\label{fig:mia_fid_tpr_fpr}
\end{figure}

\tightparagraph{Attack results vs.~FID.} To evaluate our generative models, we use the standard Fréchet Inception Distance (FID)~\cite{heusel2017gans}, where lower scores indicate higher quality.
%
Our previous CIFAR-10 results used models that achieved the best FID (on average 3.5) based on early stopping. 
Here we evaluate models over the course of training in \Cref{fig:mia_fid_tpr_fpr}. We compute the attack success rate as a function of FID, and
we find that as the quality
of the diffusion model increases so too does the privacy leakage.
These results are concerning because they suggest that stronger diffusion models of the future may be even less private.

\subsection{Inpainting Attacks} 
\label{ssec: cifar10_inpaint}

Having performed untargeted extraction on CIFAR-10 models, we now construct
a targeted version of our attack.
As mentioned earlier, performing a targeted attack is complicated by the fact that these models do not support textual prompting.
We instead provide guidance by performing a form of attribute inference attack \cite{jayaraman2022attribute, yeom2018privacy, zhang2020secret}
that we call an ``inpainting attack''.
Given an image, we first mask out a portion of this image; our attack objective is to recover the masked region.
We then run this attack on both training and testing images,
and compare the attack efficacy on each.
Specifically, for an image $x$, we mask some fraction of pixels to create a masked image $x_m$, 
and then use the trained model to 
reconstruct
the image as $x_{rec}$. 
The exact algorithm we use for inpainting is given in Lugmayr \emph{et al.}~\cite{lugmayr2022repaint}.


Because diffusion model inpainting is stochastic 
(it depends on the random sample $\epsilon \sim \mathcal{N}(0,I)$),
we create a set of inpainted images $X_{rec}=\{x^1_{rec}, x^2_{rec}, \ldots, x^n_{rec}\}$, where we set $n=5{,}000$.
For each $x_{rec}\in X_{rec}$, we compute the diffusion model's loss on this sample (at timestep 100) divided by a shadow model's loss that was not trained on the sample.
We then use this score to identify the highest-scoring reconstructions $x_{rec}\in X_{rec}$.



\begin{figure}
    \centering
    \includegraphics[width=\linewidth]{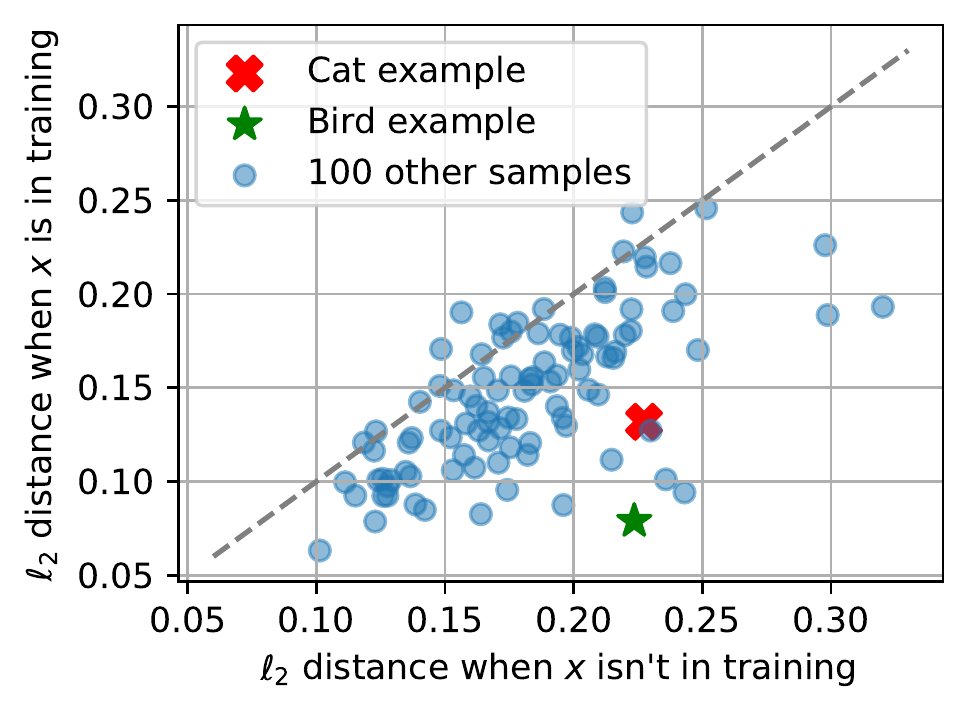}
    \caption{Evaluating inpainting attacks on $100$ CIFAR-10 examples,
    measuring the $\ell_2$ distance between images and their inpainted reconstructions when we mask out the left half of the image for 100 randomly selected images. 
    We also plot the $\ell_2$ distances for the bird and cat examples shown in \Cref{fig:example_inpaint_attack}. When an adversary has partial knowledge of an image, inpainting attacks work far better than typical data extraction.}
    \label{fig:inpaint_attack_distances}
\end{figure}

\begin{figure}[t]
  \centering
\begin{subfigure}{0.99\columnwidth}
\centering
    \includegraphics[width=0.95\linewidth]{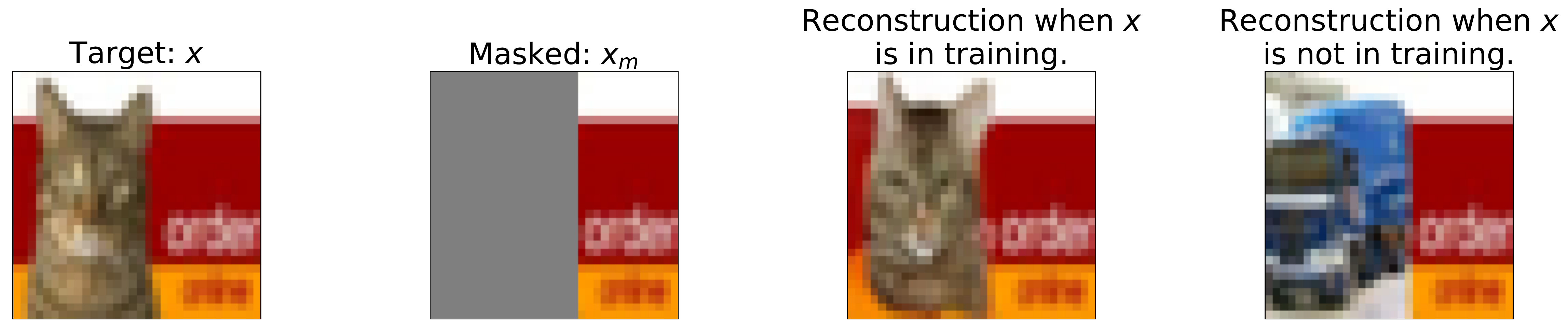}
\end{subfigure}
\begin{subfigure}{0.99\columnwidth}
\centering
    \includegraphics[width=0.95\linewidth]{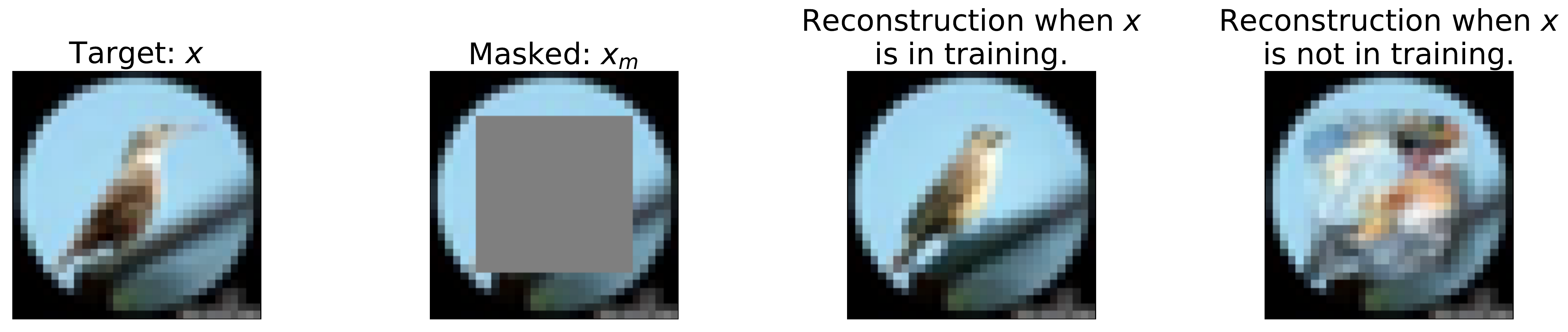}
\end{subfigure}%
\caption{Inpainting-based reconstruction attack on CIFAR-10.
Given an image from CIFAR-10 (first column), 
we randomly mask half of the image (second column),
and then inpaint the image for a model which contained this image in the training set (third column)
versus inpainting the image for a model which did not contain this image in the training set (fourth column).}
\label{fig:example_inpaint_attack}
\end{figure}

\paragraph{Results.}
Our specific attack masks out the left half of an image and applies the diffusion model on the right half of the image to inpaint the rest. 
We repeat this process 5000 times and take the top-10 scoring reconstructions using a membership inference attack. 
We repeat this attack for 100  images using diffusion models that are trained with and without the images. \Cref{fig:inpaint_attack_distances} compares the average distance between the sample and the ten highest scoring inpainted samples.
This allows us to show our inpainting attacks have succeed: the reconstruction loss is substantially better in terms of $\ell_2$ distance when the image is in the training set than when not.
\Cref{fig:example_inpaint_attack} also shows qualitative examples of this attack.
The highest-scoring reconstruction looks visually similar to the target image when the target is in training and does not resemble the target when it is not in training.
Overall, these results show that an adversary who has partial knowledge of an image can substantially improve their extraction results.
We conduct a more thorough analysis of inpainting attacks in \Cref{sec: cifar10_inpaint_more}. 

\section{Comparing Diffusion Models to GANs}

Are diffusion models more or less private than competing generative modeling approaches?
In this section we take a first look at this question by comparing diffusion models to Generative Adversarial Networks (GANs) \cite{goodfellow2020generative,salimans2016improved,radford2015unsupervised}, an approach that has held
the state-of-the-art results for image generation for nearly a decade.

Unlike diffusion models that are explicitly trained to memorize and reconstruct their training datasets, GANs are not.
Instead, GANs consist of two competing neural networks: a generator and a discriminator. Similar to diffusion models, the generator receives random noise as input, but unlike a diffusion model, it must convert this noise to a valid image in a single forward pass.
To train a GAN, the discriminator is trained to predict if an image comes from the generator or not, and the generator is trained to fool the discriminator.
As a result, GANs differ from diffusion models in that their generators are only trained using \emph{indirect} information about the training data (i.e., using gradients from the discriminator) because they never receive training data as input, whereas diffusion models are explicitly trained to reconstruct the training set.

\paragraph{Membership inference attacks.} We first propose a privacy attack methodology for GANs.\footnote{While existing privacy attacks exist for GANs, they were proposed before the latest advancements in privacy attack techniques,  requiring us to develop our own methods which out-perform prior work.} We initially focus on membership inference attacks, where following Balle \textit{et al.} \cite{balle2022reconstructing}, we assume access to both the discriminator and generator. 
We perform membership inference using the loss threshold \cite{yeom2018privacy} and LiRA \cite{carlini2022membership} attacks, where we use the discriminator's loss as the metric. To perform LiRA, we follow a similar methodology as Section~\ref{sec:cifar10} and train 256 individual GAN models each on a random $50\%$ split of the CIFAR-10 training dataset but otherwise leave training hyperparameters unchanged.

We study three GAN architectures, all implemented using the StudioGAN framework \cite{kang2022StudioGAN}: BigGAN~\cite{brock2018large}, MHGAN~\cite{turner2019metropolis}, and StyleGAN~\cite{karras2019style}. 
Figure~\ref{fig:discriminator_lira} shows the membership inference results. Overall, diffusion models have higher membership inference leakage, e.g., diffusion models had $50\%$ TPR at a FPR of $0.1\%$ as compared to $<30\%$ TPR for GANs. This suggests that diffusion models are less private than GANs for membership inference attacks under default training settings, even when the GAN attack is strengthened due to having access to the discriminator (which would be unlikely in practice, as only the generator is necessary to create new images).

\ifarxiv
\begin{figure*}[t]
  \centering
  \begin{subfigure}{0.3\textwidth}
\centering
    \includegraphics[width=0.95\linewidth]{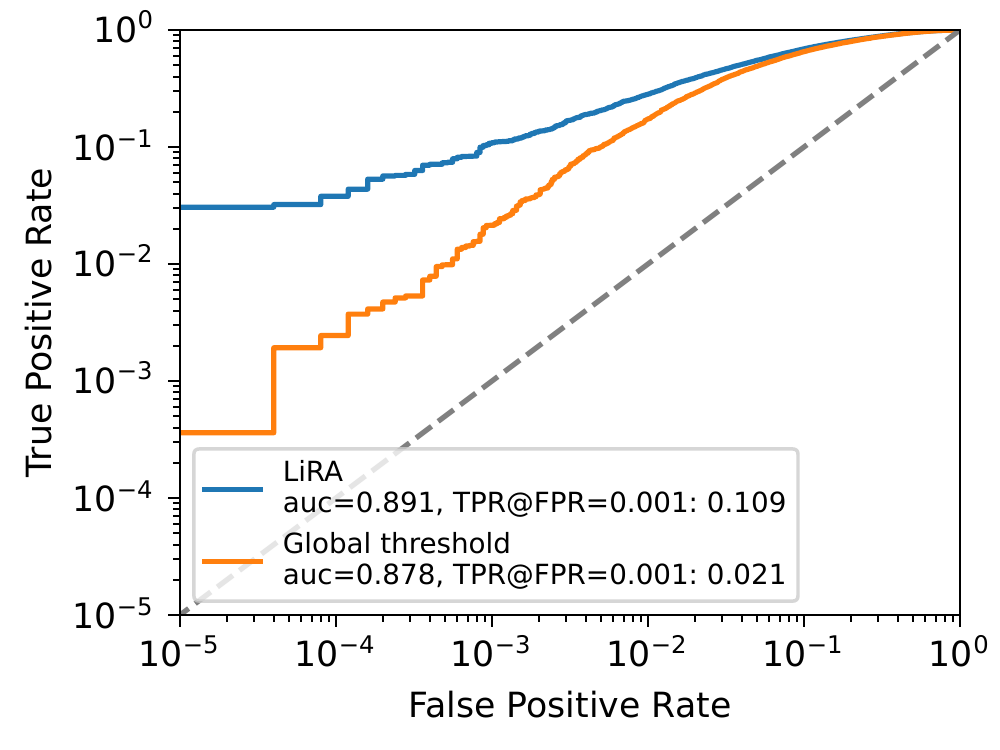}
        \caption{StyleGAN FID avg = 3.7}
    \label{fig:stylegan_lira}
\end{subfigure}%
\begin{subfigure}{0.3\textwidth}
\centering
    \includegraphics[width=0.95\linewidth]{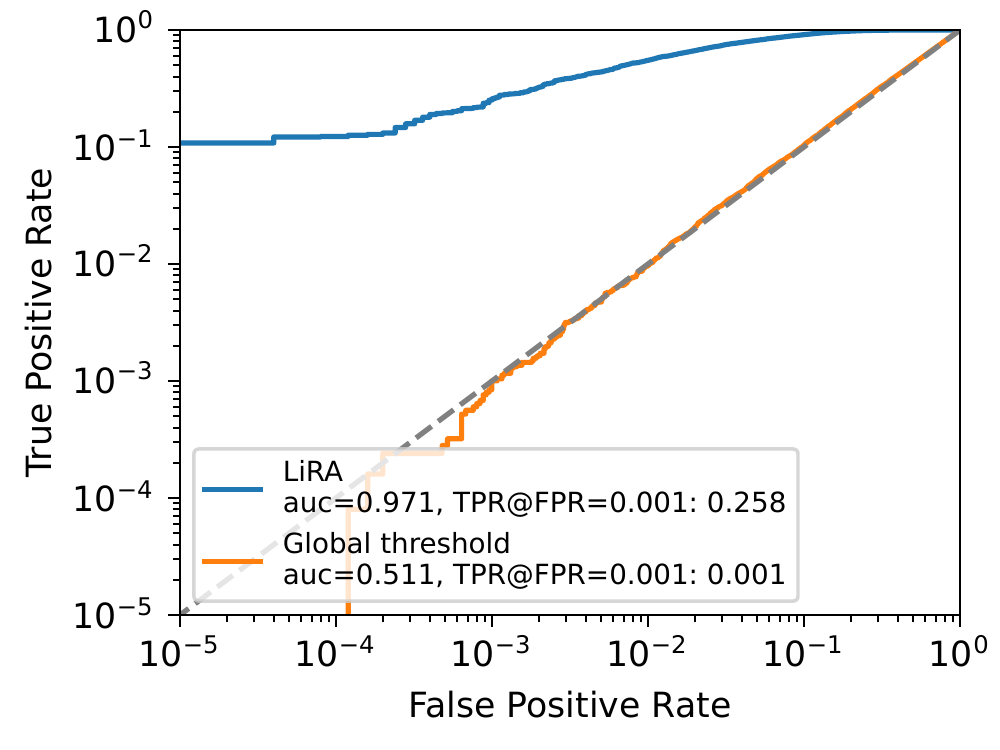}
        \caption{MHGAN FID avg = 7.9}
    \label{fig:mhgan_fprtpr}
\end{subfigure}%
\begin{subfigure}{0.3\textwidth}
\centering
    \includegraphics[width=0.95\linewidth]{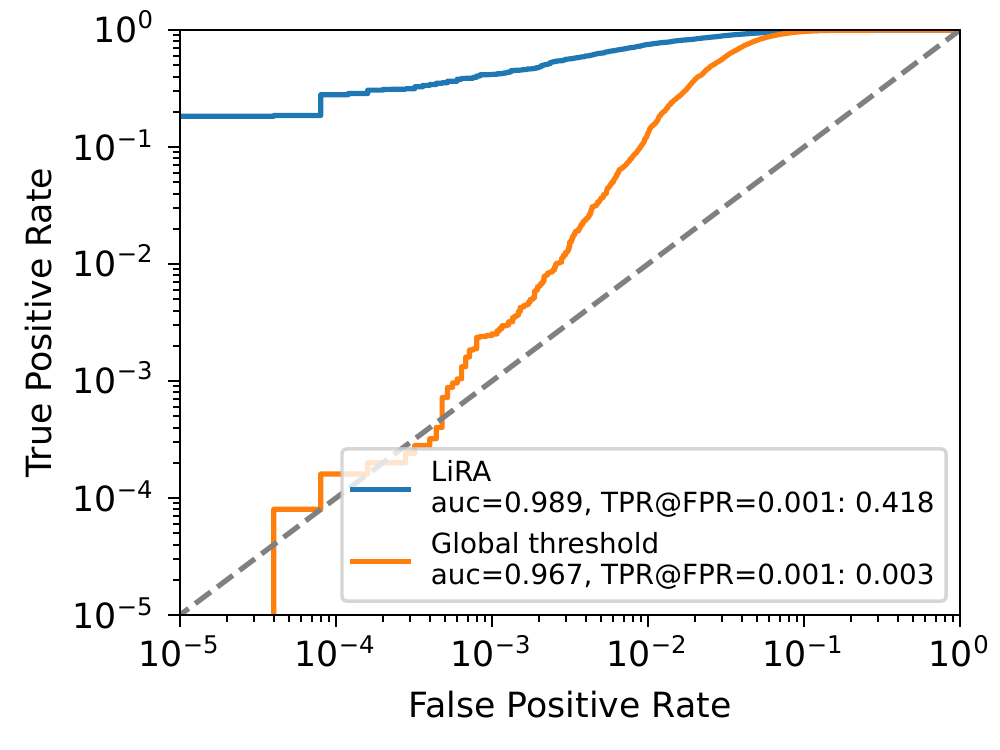}
        \caption{BigGAN FID avg = 7.7}
    \label{fig:biggan_lira}
\end{subfigure}%
\caption{Membership inference results on GAN models using the loss threshold and LiRA attacks on the discriminator. Overall, GANs are significantly more private than diffusion models under default training configurations.}
\label{fig:discriminator_lira}
\end{figure*}

\begin{figure*}[t]
    \begin{subfigure}{\textwidth}
\centering
    \includegraphics[width=0.95\linewidth]{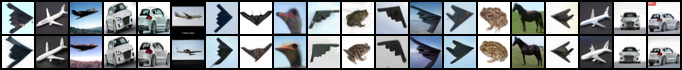}
        \caption{StyleGAN }
    \label{fig:stylegan-goodones}
\end{subfigure}%

\begin{subfigure}{\textwidth}
\centering
    \includegraphics[width=0.95\linewidth]{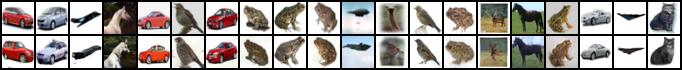}
        \caption{MHGAN }
    \label{fig:mhgan_goodones}
\end{subfigure}%

\begin{subfigure}{\textwidth}
\centering
    \includegraphics[width=0.95\linewidth]{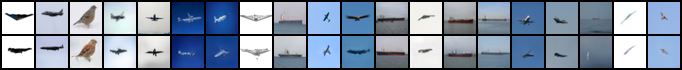}
        \caption{BigGAN }
    \label{fig:biggan_goodones}
\end{subfigure}%
    \vspace{-0.1cm}
    \caption{Selected training examples we extract from three GANs trained on CIFAR-10 for different architectures.
    \textbf{Top} row: generated output from a diffusion model.
    \textbf{Bottom} row: nearest ($\ell_2)$ example from the training dataset.
    Figure~\ref{fig:gan_cifar_all_extracted} in the Appendix contains all unique extracted images.}
    \label{fig:gan_extractions}
\end{figure*}
\fi

\begin{table}[t]
    \centering
    \footnotesize
    \begin{tabular}{@{} llrr @{}} 
        \toprule
        \multicolumn{2}{l}{\textbf{Architecture}}& \textbf{Images Extracted} & \textbf{FID} \\
        \midrule 
        
        \multirow{5}{*}{\rotatebox[]{0}{\textbf{GANs}}}
        &StyleGAN-ADA \cite{karras2020styleganAda} & \textbf{150} & \textbf{2.9} \\
        &DiffBigGAN \cite{zhao2020diffAugGAN} & 57 & 4.6 \\
        &E2GAN \cite{tian2020e2gan} & 95 & 11.3 \\
        &NDA \cite{sinha2021nda} & 70 & 12.6 \\
        &WGAN-ALP \cite{terjek2019wganALP} & 49 & 13.0 \\ 
        \midrule
        \multirow{2}{*}{\rotatebox[]{0}{\textbf{DDPMs}}}
        &OpenAI-DDPM \cite{nichol2021improved} & \textbf{301} & \textbf{2.9} \\
        &DDPM \cite{ho2020denoising} & 232 & 3.2 \\
        \bottomrule
    \end{tabular}
    \caption{The number of training images that we extract from different off-the-shelf pretrained generative models out of 1 million unconditional generations. We show GAN models sorted by FID (lower is better) on the top and diffusion models on the bottom. Overall, we find that diffusion models memorize more than GAN models. Moreover, better generative models (lower FID) tend to memorize more data.}
        \label{tab:offshelfGans}
\end{table}

\tightparagraph{Data extraction results.} 
We next turn our attention away from measuring worst-case privacy risk and
focus our attention on more practical black-box extraction attacks.
We follow the same procedure as Section~\ref{ssec:cifar10_extract}, where we generate $2^{20}$ images from each model architecture and identify those that are near-copies of the training data using the same similarity function as before. 
Again we only consider non-duplicated CIFAR-10 training images in our counting. 
For this experiment, instead of using models we train ourselves (something that was necessary to run LiRA), we study five off-the-shelf pre-trained GANs: WGAN-ALP~\cite{terjek2019wganALP}, E2GAN~\cite{tian2020e2gan}, NDA~\cite{sinha2021nda}, DiffBigGAN~\cite{zhao2020diffAugGAN}, and StyleGAN-ADA~\cite{karras2020styleganAda}. We also evaluate two off-the-shelf DDPM diffusion model released by Ho \textit{et al.}~\cite{ho2020denoising} and Nichol \emph{et al.}~\cite{nichol2021improved}. Note that all of these pre-trained models are trained by the original authors to maximize utility on the entire CIFAR-10 dataset rather than a random 50\% split as in our prior models trained for MIA.

Table~\ref{tab:offshelfGans} shows the number of extracted images for each model and their corresponding FID. Overall, we find that diffusion models memorize more data than GANs, even when the GANs reach similar performance, e.g., the best DDPM model memorizes $2\times$ more than StyleGAN-ADA but reaches the same FID. Moreover, generative models (both GANs and diffusion models) tend to memorize more data as their quality (FID) improves, e.g., StyleGAN-ADA memorizes $3\times$ more images than the weakest GANs. 

Using the GANs we trained ourselves, we show examples of the near-copy generations in Figure~\ref{fig:gan_extractions} for the three GANs that we trained ourselves, and Figure~\ref{fig:gan_extractions_3} in the Appendix shows every sample that we extract for those models. 
The Appendix also contains near-copy generations from the five off-the-shelf GANs.
Overall, these results further reinforce the conclusion that diffusion models are less private than GAN models. 

We also surprisingly find that diffusion models and GANs memorize many of the same images. In particular, despite the fact that our diffusion model memorizes 1280 images and a StyleGAN model we train on half of the dataset memorizes 361 images, 
we find that \emph{244 unique images are memorized in common}. 
If images were memorized uniformly at random, we should expect on average $10$ images would be memorized by both, giving exceptionally strong evidence that some images $(p < 10^{-261})$
are inherently less private than others. Understanding why this phenomenon occurs is a fruitful direction for future work.

\section{Defenses and Recommendations}

Given the degree to which diffusion models memorize and regenerate training examples, in this section we explore various defenses and practical strategies that may help to reduce and audit model memorization.

\subsection{Deduplicating Training Data}

In Section~\ref{ssec:stablediffusion}, we showed that many examples that are easy to extract are duplicated many times (e.g., $>100$) in the training data. Similar results have been shown for language models for text~\cite{carlini2021extracting,kandpal2022large} and data deduplication has been shown to be an effective mitigation against memorization for those models~\cite{lee2021deduplicating,kandpal2022deduplicating}. In the image domain, simple deduplication is common, where images with identical URLs and captions are removed, but most datasets do not compute other inter-image similarity metrics such as $\ell_2$ distance or CLIP similarity. We thus encourage practitioners to deduplicate future datasets using these more advanced notions of duplication.

Unfortunately, deduplication is not a perfect solution.
To better understand the effectiveness of data deduplication, we deduplicate CIFAR-10 and re-train a diffusion model on this modified dataset. 
We compute image similarity using the \texttt{imagededup} tool and deduplicate any images that have a similarity above $>0.85$. This removes $5{,}275$ examples from the $50{,}000$ total examples in CIFAR-10. 
We repeat the same generation procedure as Section~\ref{ssec:cifar10_extract}, where we generate $2^{20}$ images from the model and count how many examples are regenerated from the training set. The model trained on the deduplicated data regenerates $986$ examples, as compared to $1280$ for the original model. While not a substantial drop, these results show that deduplication can mitigate memorization. Moreover, we also expect that deduplication will be much more effective for models trained on larger-scale datasets (e.g., Stable Diffusion), as we observed a much stronger correlation between data extraction and duplication rates for those models.

\subsection{Differentially-Private Training}

The gold standard technique to defend against privacy attacks is by training with differential privacy (DP) guarantees~\cite{dwork2006calibrating,dwork2008differential}. Diffusion models can be trained with differentially-private stochastic gradient descent (DP-SGD)~\cite{abadi2016deep}, where the model's gradients are clipped and noised to prevent the model from leaking substantial information about the presence of any individual image in the dataset. 
Applying DP-SGD induces a trade-off between privacy and utility, and recent work shows that DP-SGD can be applied to small-scale diffusion models without substantial performance degradation~\cite{dockhorn2022differentially}.

Unfortunately, we applied DP-SGD to our diffusion model codebase and found that it caused the training on CIFAR-10 to consistently diverge, even at high values for $\epsilon$ (the privacy budget, around 50). In fact, even applying a non-trivial gradient clipping or noising on their own (both are required in DP-SGD) caused the training to fail. We leave a further investigation of these failures to future work, and we believe that new advances in DP-SGD and privacy-preserving training techniques may be required to train diffusion models in privacy-sensitive settings.

\begin{figure}
    \centering
    \includegraphics[width=0.8\linewidth]{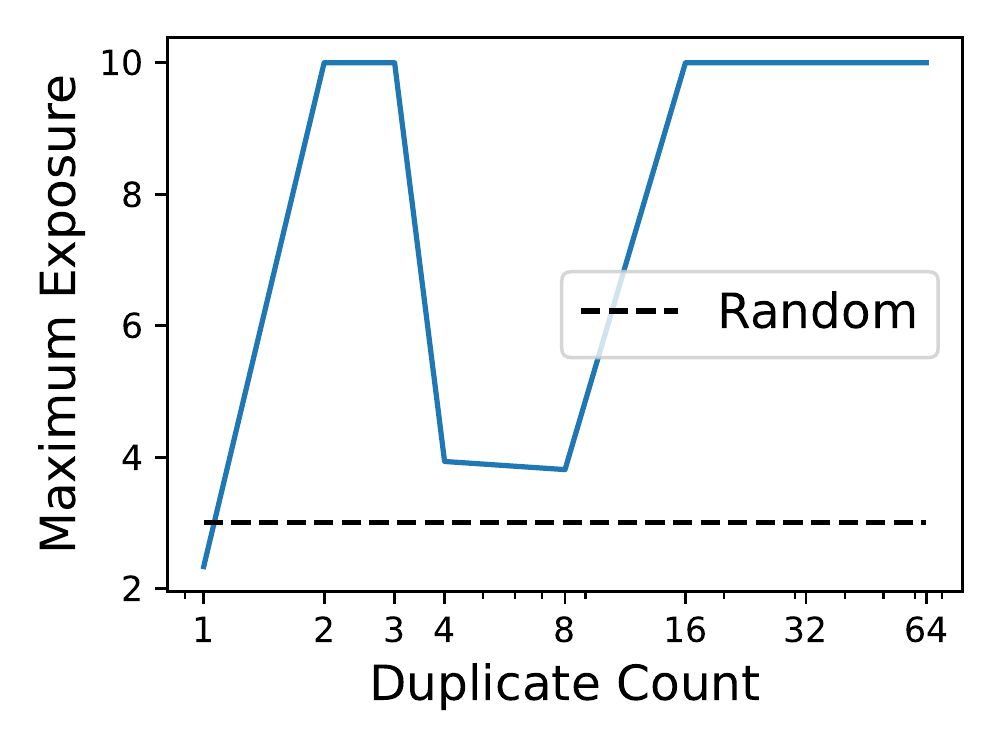}
    \vspace{-0.4cm}
    \caption{Canary \emph{exposure} (a measure of non-privacy) as a function of duplicate count. Inserting a canary twice is sufficient to reach maximum exposure.}
    \label{fig:canaries}
\end{figure}

\subsection{Auditing with Canaries}


In addition to implementing defenses, it is important for practitioners to empirically audit their models to determine how vulnerable they are in practice~\cite{jagielski2020auditing}. 
Our attacks above represent one method to evaluate model privacy. Nevertheless, our attacks are expensive, e.g., our membership inference results require training many shadow models, and thus lighter weight alternatives may be desired.

One such alternative is to insert canary examples into the training set, a common approach to evaluate memorization in language models~\cite{carlini2019secret}. Here, one creates a large ``pool'' of \emph{canaries}, e.g., by randomly generating noise images, and inserts a subset of the canaries into the training set. After training, one computes the \emph{exposure} of the canaries, which roughly measures how many bits were learned about the inserted canaries as compared to the larger pool of not inserted canaries. This loss-based metric only requires training one model and can also be designed in a worst-case way (e.g., adversarial worst-case images could be used).

To evaluate exposure for diffusion models, we generate canaries consisting of uniformly generated noise. We then duplicate the canaries in the training set at different rates and measure the maximum exposure. Figure~\ref{fig:canaries} shows the results. Here, the maximum exposure is 10, and some canaries reach this exposure after being inserted only twice. The exposure is not strictly increasing with duplicate count, which may be a result of some canaries being ``harder'' than others, and, ultimately, random canaries we generate may not be the most effective canaries to use to test memorization for diffusion models.

\section{Related Work}\label{sec:related-work}

\paragraph{Memorization in language models.} Numerous past works study memorization in generative models across different domains, architectures, and threat models. One area of recent interest is memorization in language models for text, where past work shows that adversaries can extract training samples using two-step attack techniques that resemble our approach~\cite{carlini2021extracting,lee2021deduplicating,kandpal2022deduplicating,kandpal2022large}. Our work differs from these past results because we focus on the image domain and also use more semantic notions of data regeneration (e.g., using CLIP scores) as opposed to focusing on exact verbatim repetition (although recent language modeling work has begun to explore approximate memorization as well \cite{ippolito2022preventing}).

\paragraph{Memorization in image generation.} Aside from language modeling, past work also analyzes memorization in image generation, mainly from the perspective of generalization in GANs (i.e., the novelty of model generations). For instance, numerous metrics exist to measure similarity with the training data~\cite{heusel2017gans,arora2018do}, the extent of mode collapse~\cite{salimans2016improved,che2016mode}, and the impact of individual training samples~\cite{balaji2019entropic,van2021memorization}. Moreover, other work provides insights into when and why GANs may replicate training examples~\cite{nagarajan2018theoretical,feng2021gans}, as well as how to mitigate such effects~\cite{nagarajan2018theoretical}. Our work extends these lines of inquiry to conditional diffusion models, where we measure novelty by computing how frequently models regenerate training instances when provided with textual prompts.

Recent and concurrent work also studies privacy in image generation for both GANs~\cite{tinsley2021face} and diffusion models~\cite{somepalli2022diffusion,wu2022membership,hu2023membership}.
Tinsley \emph{et al.}~\cite{tinsley2021face} show that StyleGAN can generate individuals' faces, and Somepalli \emph{et al.}~\cite{somepalli2022diffusion} show that Stable Diffusion can output semantically similar images to its training set. Compared to these works, we identify privacy vulnerabilities in a wider range of systems (e.g., Imagen and CIFAR models) and threat models (e.g., membership inference attacks).

\section{Discussion and Conclusion}

State-of-the-art diffusion models memorize and regenerate individual training images, 
allowing adversaries to launch training data extraction attacks.
By training our own models we find that increasing utility can degrade privacy,
and simple defenses such as deduplication are insufficient to completely address the memorization challenge.
We see that state-of-the-art diffusion models memorize $2\times$ more than comparable GANs,
and more useful diffusion models memorize more than weaker diffusion models.
This suggests that the vulnerability of generative image models may grow over time.
Going forward, our work raises questions around the memorization and generalization capabilities of diffusion models.

\paragraph{Questions of generalization.} 
Do large-scale models work by generating novel output, or do they just copy and interpolate between individual training examples?
If our extraction attacks had failed, 
it may have refuted the hypothesis that models copy and interpolate training data;
but because our attacks succeed, this question remains open.
%
Given that different models memorize varying amounts of data,
we hope future work will explore how
diffusion models copy from their training datasets.

Our work also highlights the difficulty in defining \emph{memorization}.
While we have found extensive memorization with a simple $\ell_2$-based measurement,
a more comprehensive analysis will be necessary to accurately capture more 
nuanced definitions of memorization that allow for more
human-aligned notions of data copying.


\paragraph{Practical consequences.} 
We raise four practical consequences for those who train and deploy diffusion models.
First, while not a perfect defense, we recommend deduplicating training datasets and minimizing over-training.
Second, we suggest using our attack---or other auditing techniques---to estimate the privacy risk of trained models. 
Third, once practical privacy-preserving techniques become possible, we recommend their use whenever possible.
Finally, we hope our work will temper the heuristic
privacy expectations that have come to be associated with diffusion model outputs:
synthetic data does not give privacy for free \cite{Chambon2,chambon,Rouzrokh,Ali,pinaya}.

\begin{table*}[]
    \centering
    \begin{tabular}{@{}l|ccc|ccc|ccc|@{}}
    
        \multicolumn{1}{c}{}& \multicolumn{1}{c}{NC} &
        MN &
        \multicolumn{1}{c}{JH} &
        \multicolumn{1}{c}{MJ} &
        FT &
        \multicolumn{1}{c}{VS} &
        \multicolumn{1}{c}{BB} &
        DI &
        \multicolumn{1}{c}{EW} \\
        \toprule
        Conceived Project & X & & X & & & X &&&X\\
        Formalized Memorization Definition & X & X & X & X & X & & X && \\
        Experimented with Stable Diffusion & X & X & &&&&&&\\
        Experimented with Imagen &  & X & &&&&&&\\
        Experimented with CIFAR-10 Diffusion & X &  & X &&&&&&\\
        Experimented with GANs &  & X &  &&X&X&&&\\
        Experimented with Defenses & X & X &  &X&&&&&\\
        Prepared Figures & X & X & X &X&&X&&X&X\\
        Analyzed Data & X & X & X & X & X & X & & &  \\
        Wrote Paper & X&X&X&X&X&X&X&X&X\\
        Managed the Project &X & & & & & & & &  \\
        \bottomrule
    \end{tabular}
    \caption{Contributions of each author in the paper.}
    \label{tab:my_label}
\end{table*}

On the whole, 
our work contributes to a growing body of literature that raises questions regarding the legal, ethical, and privacy issues that arise from training on web-scraped public data~\cite{brown2022does,somepalli2022diffusion,tramer2022considerations,wallace2020gpt2}. 
%
Researchers and practitioners should be wary of training on uncurated public data without first taking steps to understand the underlying ethics and privacy implications.

\section*{Contributions}

\begin{itemize}[itemsep=0pt]
    \item Nicholas, Jamie, Vikash, and Eric each independently proposed the
    problem statement of extracting training data from diffusion models.
    \item Nicholas, Eric, and Florian performed preliminary experiments to identify
    cases of data extraction in diffusion models.
    \item Milad performed most of the experiments on Stable Diffusion and Imagen,
    and Nicholas counted duplicates in the LAION training dataset;
    each wrote the corresponding sections of the paper.
    \item Jamie performed the membership inference attacks and inpainting attacks on CIFAR-10 diffusion models,
    and Nicholas performed the diffusion extraction experiments;
    each wrote the corresponding sections of the paper. 
    \item Matthew ran experiments for canary memorization and wrote the corresponding section of the paper.
    \item Florian and Vikash performed preliminary experiments on memorization in GANs, 
    and Milad and Vikash ran the experiments included in the paper.
    \item Milad ran the membership inference experiments on GANs. 
    \item Vikash ran extraction experiments on pretrained GANs.
    \item Daphne and Florian improved figure clarity and presentation.
    \item Daphne, Borja, and Eric  edited the paper and contributed to paper framing.
    \item Nicholas organized the project and wrote the initial paper draft.
\end{itemize}

\section*{Acknowledgements and Conflicts of Interest}

The authors are grateful to Tom Goldstein, Olivia Wiles, Katherine Lee, Austin Tarango, Ian Wilbur, Jeff Dean, Andreas Terzis, Robin Rombach, and Andreas Blattmann for comments on early drafts of this paper.

Nicholas, Milad, Matthew, and Daphne are employed at Google, 
and Jamie and Borja are employed at DeepMind,
companies that both train large machine learning models (including diffusion models) on both public and private datasets.

Eric Wallace is supported by the Apple Scholars in AI/ML Fellowship.

\bibliographystyle{plain}
\bibliography{main}

\newpage
\appendix
\clearpage
\onecolumn

\section{Collected Details for Figures}

\begin{table}[h!]
    \centering
    \setlength{\tabcolsep}{10pt}
    \caption{Catalog of figures containing qualitative examples.}
    \label{tab:figures}
    \begin{tabular}{@{}l l  l l l@{}}
    \toprule
    Figure \# & Model & Dataset & Who trained it? &Sampling strategy \\
    \midrule
\Cref{fig:teaser} & Stable Diffusion & LAION & Stability AI & PLMS \\
\Cref{fig:obama} & Stable Diffusion & LAION & Stability AI & PLMS \\
\Cref{fig:sd_14_extractions_sample} & Stable Diffusion & LAION & Stability AI & PLMS \\
\Cref{fig:badl2} & Uncond Diffusion & CIFAR-10 & Ours & DDIM \\
\Cref{fig:lossdistribution} & Uncond Diffusion & CIFAR-10 & Ours & DDIM \\
\Cref{fig:extracted_selected} & Uncond Diffusion & CIFAR-10 & Ours & DDIM \\
\Cref{fig:inpaint_attack_distances} & Uncond Diffusion & CIFAR-10 & Ours & Inpainting \\
\Cref{fig:example_inpaint_attack} & Uncond Diffusion & CIFAR-10 & Ours & Inpainting \\
\Cref{fig:gan_extractions} & StyleGAN, MHGAN, BigGAN & CIFAR-10 & Ours & GAN default \\
\Cref{fig:cifar_all_extracted} & Uncond Diffusion & CIFAR-10 & Ours & DDIM \\
\Cref{fig:inliers} & Uncond Diffusion & CIFAR-10 & Ours & DDIM \\
\Cref{fig:inpainting_rec_attack_train_bird} & Uncond Diffusion & CIFAR-10 & Ours & Inpainting \\
\Cref{fig:inpainting_rec_attack_test_bird} & Uncond Diffusion & CIFAR-10 & Ours & Inpainting \\
\Cref{fig:gan_extractions_3} & Several different GANs & CIFAR-10 & Original paper authors & GAN default\\
    \bottomrule
    \end{tabular}
\end{table}

\newpage

\section{All CIFAR-10 Memorized Images}

\begin{figure*}[h!]
    \centering
    \includegraphics[scale=.25]{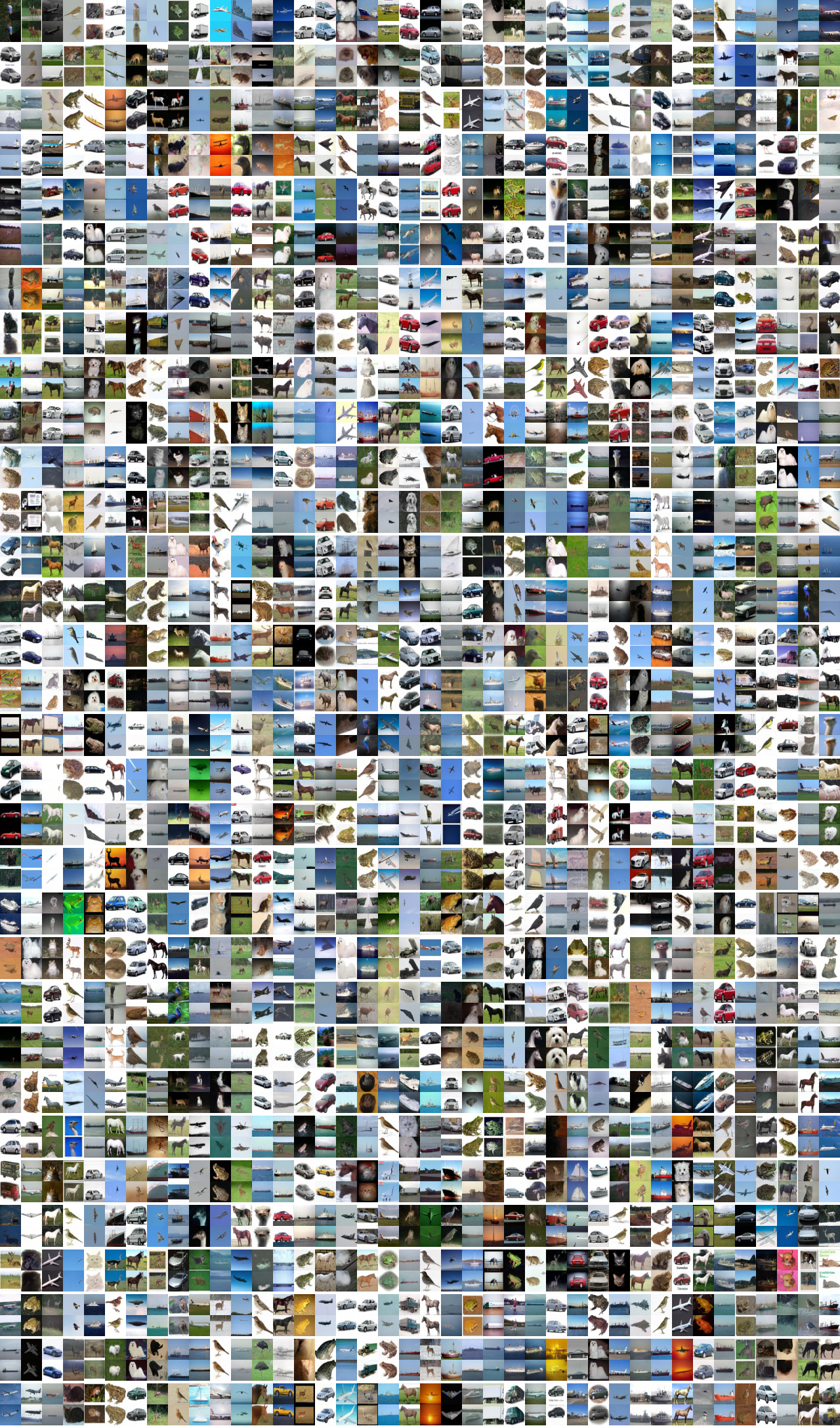}
    \caption{All 1280 images we extract from diffusion models trained on CIFAR-10,
    after 1 million generations from 16 diffusion models.}
    \label{fig:cifar_all_extracted}
\end{figure*}
\newpage

\ifarxiv
\else
\begin{figure*}[t]
  \centering
  \begin{subfigure}{0.3\textwidth}
\centering
    \includegraphics[width=0.95\linewidth]{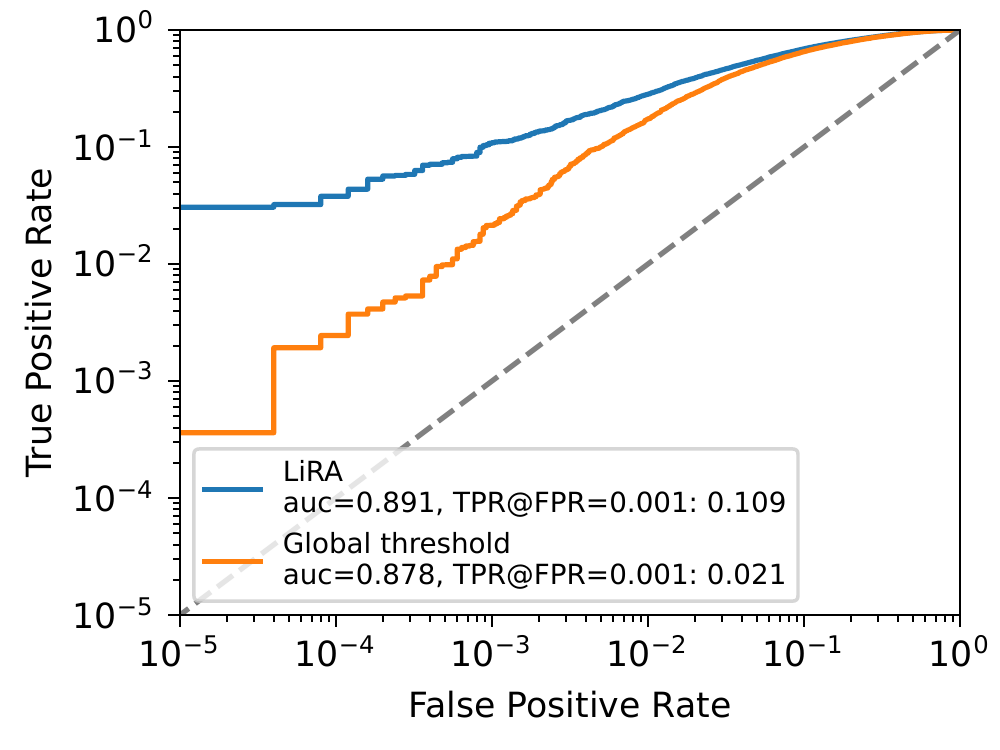}
        \caption{StyleGAN FID avg = 3.7}
    \label{fig:stylegan_lira}
\end{subfigure}%
\begin{subfigure}{0.3\textwidth}
\centering
    \includegraphics[width=0.95\linewidth]{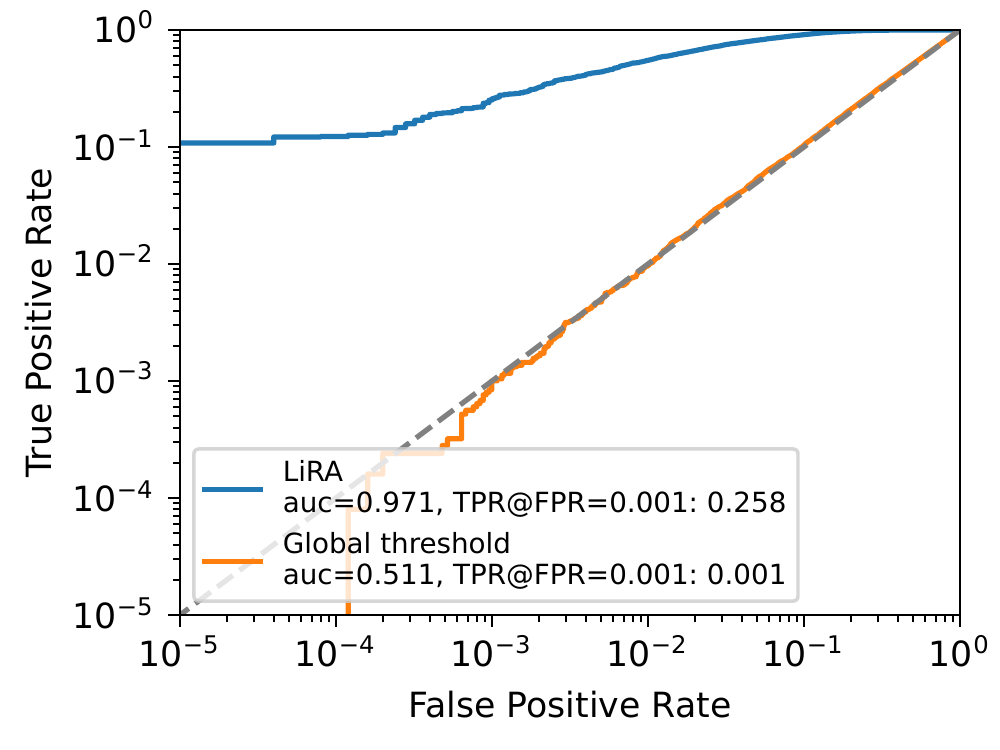}
        \caption{MHGAN FID avg = 7.9}
    \label{fig:mhgan_fprtpr}
\end{subfigure}%
\begin{subfigure}{0.3\textwidth}
\centering
    \includegraphics[width=0.95\linewidth]{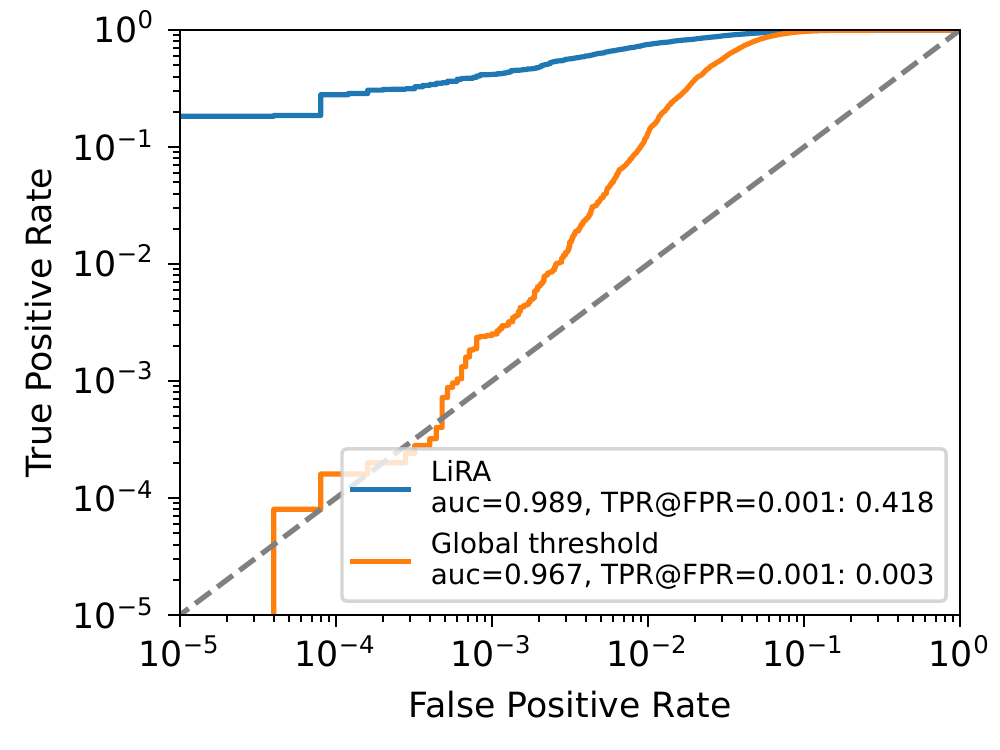}
        \caption{BigGAN FID avg = 7.7}
    \label{fig:biggan_lira}
\end{subfigure}%
\caption{Membership inference results on GAN models using the loss threshold and LiRA attacks on the discriminator. Overall, GANs are significantly more private than diffusion models under default training configurations.}
\label{fig:discriminator_lira}
\end{figure*}

\begin{figure*}[t]
    \begin{subfigure}{\textwidth}
\centering
    \includegraphics[width=0.95\linewidth]{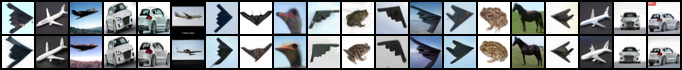}
        \caption{StyleGAN }
    \label{fig:stylegan-goodones}
\end{subfigure}%

\begin{subfigure}{\textwidth}
\centering
    \includegraphics[width=0.95\linewidth]{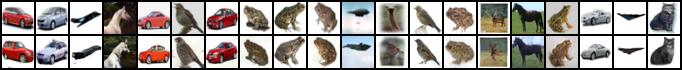}
        \caption{MHGAN }
    \label{fig:mhgan_goodones}
\end{subfigure}%

\begin{subfigure}{\textwidth}
\centering
    \includegraphics[width=0.95\linewidth]{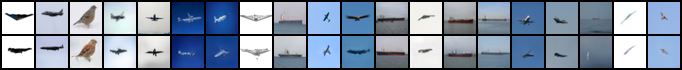}
        \caption{BigGAN }
    \label{fig:biggan_goodones}
\end{subfigure}%
    \vspace{-0.1cm}
    \caption{Selected training examples we extract from three GANs trained on CIFAR-10 for different architectures.
    \textbf{Top} row: generated output from a diffusion model.
    \textbf{Bottom} row: nearest ($\ell_2)$ example from the training dataset.
    Figure~\ref{fig:gan_cifar_all_extracted} in the Appendix contains all unique extracted images.}
    \label{fig:gan_extractions}
\end{figure*}
\fi

\section{Additional Attacks on CIFAR-10}

Here, we expand on our investigation of memorization of training data on CIFAR-10.


\subsection{Membership Inference at Different Training Steps}


\begin{figure*}[h!]
\captionsetup{width=0.9\textwidth, justification=centering}
  \centering
\begin{subfigure}[t]{0.3\textwidth}
\centering
    \includegraphics[width=0.95\linewidth]{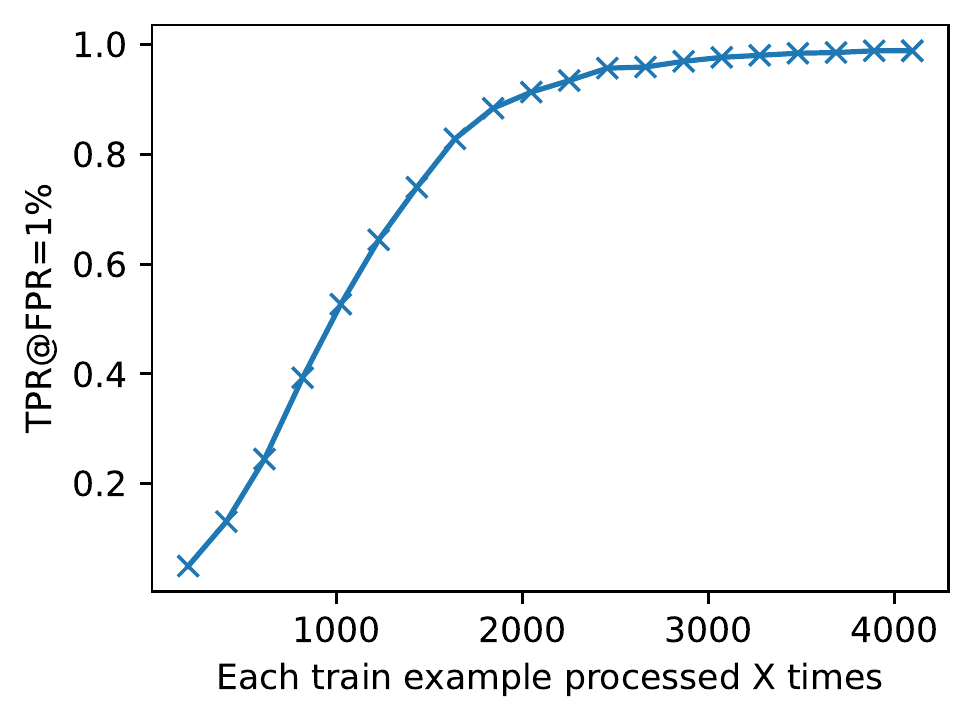}
        \caption{How membership attack success changes as a training example is processed repeatedly throughout training.}
    \label{fig:mia_cifar10_example_processed}
\end{subfigure}%
\begin{subfigure}[t]{0.3\textwidth}
\centering
    \includegraphics[width=0.95\linewidth]{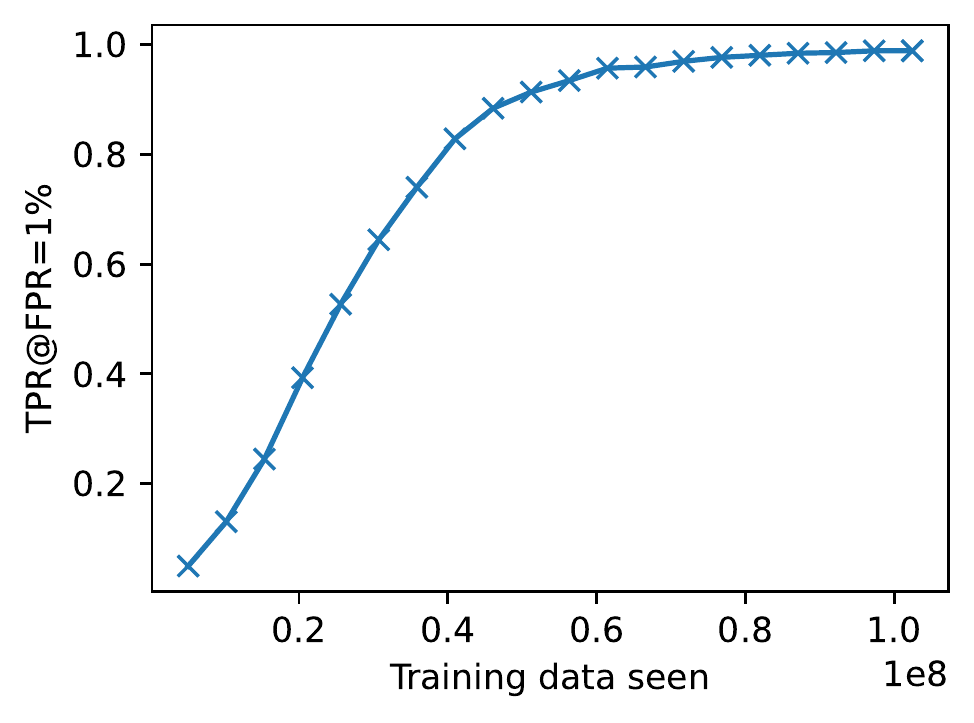}
        \caption{How membership attack success changes as more data is processed throughout training.}
         \label{fig:mia_cifar10_data_see}
\end{subfigure}%
\begin{subfigure}[t]{0.3\textwidth}
\centering
    \includegraphics[width=0.95\linewidth]{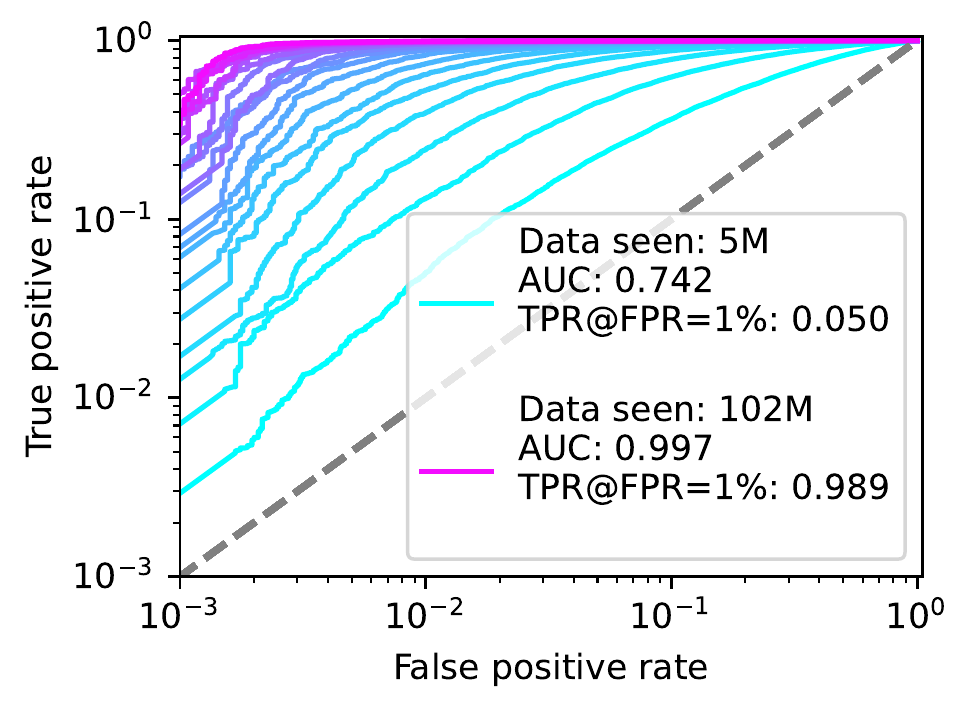}
        \caption{ROC curve for the membership attack for different training steps.}
         \label{fig:mia_cifar10_training_snapshot_roc}
\end{subfigure}%
\caption{Membership inference attacks as a function of the amount of training data processed on CIFAR-10.}
\label{fig:mia_vs_training_time_cifar10}
\end{figure*}

In \Cref{ssec:mia_vs_fid}, we implicitly investigated membership attack success as a function of the number update steps when training a diffusion model. 
We explicitly model this relationship in \Cref{fig:mia_vs_training_time_cifar10}.
First, in \Cref{fig:mia_cifar10_example_processed} we plot membership attack success as a function of the number of times that an example was processed over training.
If an example is processed more than 2000 times during training, invariably membership attacks are perfect against that example.
Second, in \Cref{fig:mia_cifar10_data_see}, we plot membership attack success as a function of the total amount of data processed during training.
Unsurprisingly, membership attack success increases as more training data is processed.
This is highlighted in \Cref{fig:mia_cifar10_training_snapshot_roc}, where we plot the membership attack ROC curve. 
At 5M training examples processed, at a FPR of 1\% the TPR is 5\%, and increases to 99\% after 102M examples are processed.
Note that this number of processed training inputs is commonly used in diffusion model training.
For example, the OpenAI CIFAR-10 diffusion model~\footnote{\url{https://github.com/openai/improved-diffusion}} is trained for 500,000 steps at a batch size of 128, meaning 64M training examples are processed. 
Even at this number of processed training examples, our membership attack has a TPR $>95\%$ at a FPR of 1\%.

\newpage
\subsection{Membership Inference with Different Augmentation Strategies}

\noindent
\begin{figure*}[h!]
  \centering
\begin{subfigure}{0.45\textwidth}
\centering
    \includegraphics[width=0.98\linewidth]{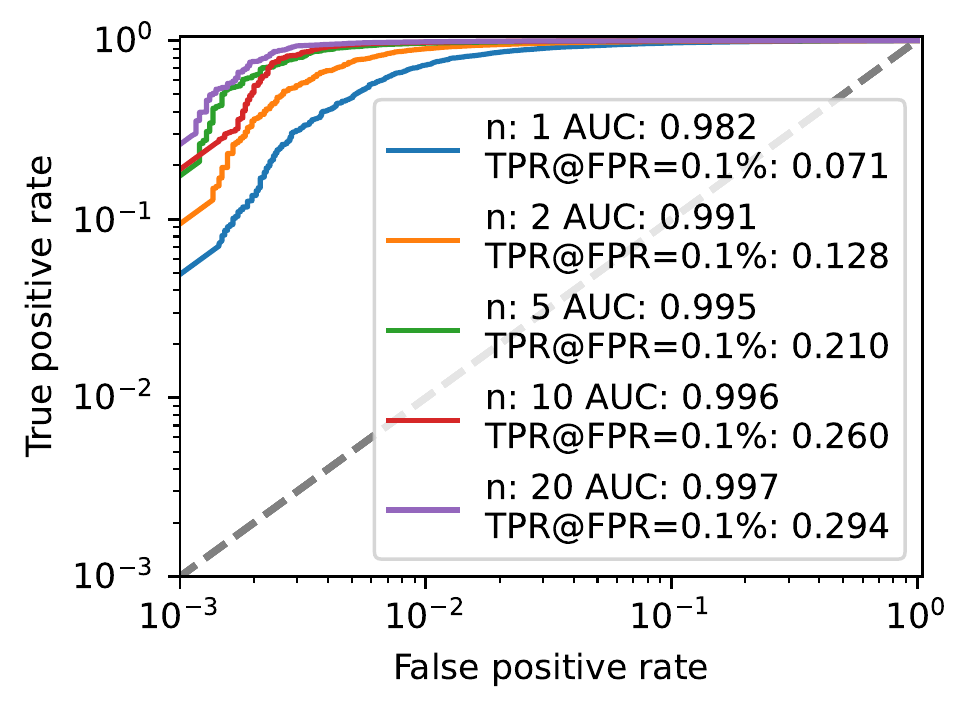}
    \caption{}
    \label{fig:mia_repeated_noise}
\end{subfigure}%
\begin{subfigure}{0.45\textwidth}
\centering
    \includegraphics[width=0.98\linewidth]{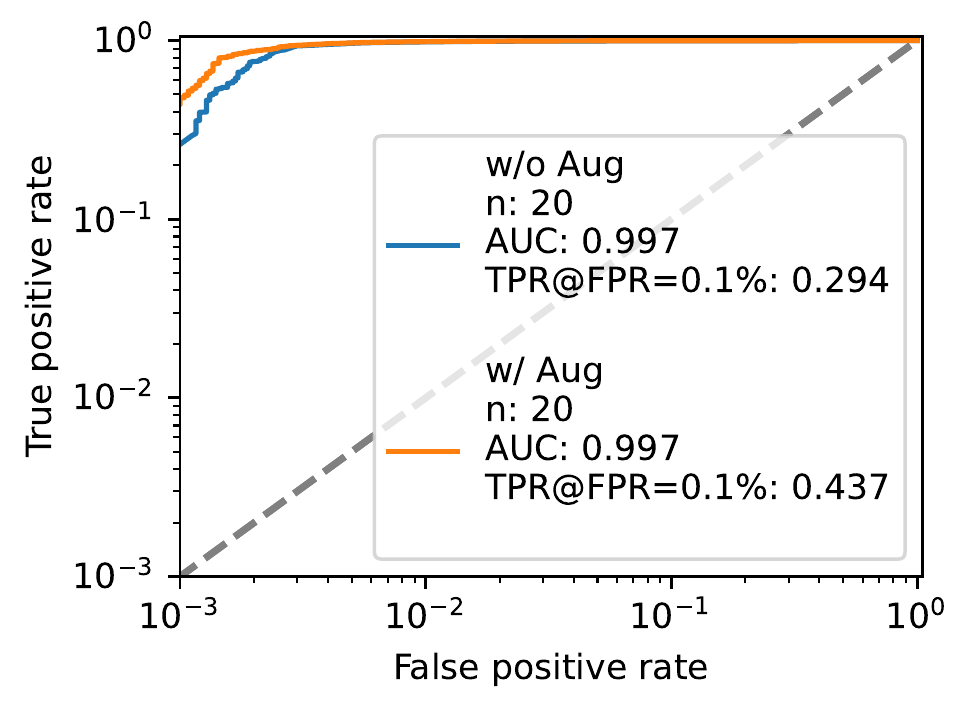}
    \caption{}
         \label{fig:mia_repeated_noise_and_flip}
\end{subfigure}%
\vspace{-0.2cm}
\caption{We can improve membership inference attack success rates on CIFAR-10 by reducing noise. In \textbf{(a)}, membership inference attacks are improved by averaging the loss over multiple noise samples in the diffusion process. In \textbf{(b)}, attacks are improved by querying on augmented versions of the candidate image.}
\label{fig:improved_mia_with_repeated_noise}
\end{figure*}

\newpage
\subsection{Membership Inference Inliers and Outliers}

\begin{figure}[h!]
    \centering
    \includegraphics[scale=.5]{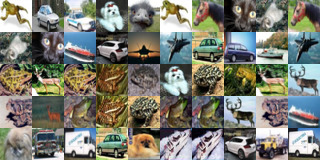}
    \hspace{1em}
    \includegraphics[scale=.5]{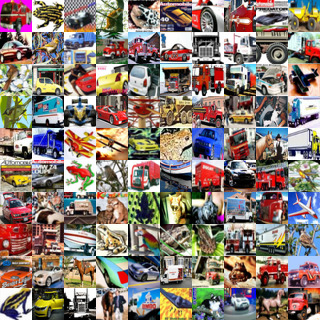}
    \caption{When performing our membership inference attack,
    the hardest-to-attack examples (left) are all duplicates in the CIFAR-10 training set,
    and the easiest-to-attack examples (right) are visually outliers from CIFAR-10 images.}
    \label{fig:inliers}
\end{figure}

\newpage

\subsection{Membership Inference on Conditional and Unconditional Models}\label{ssec:mia_cond_vs_uncond}

\begin{figure}[h!]
\captionsetup{width=0.5\textwidth, justification=centering}
  \centering
    \includegraphics[width=0.5\linewidth]{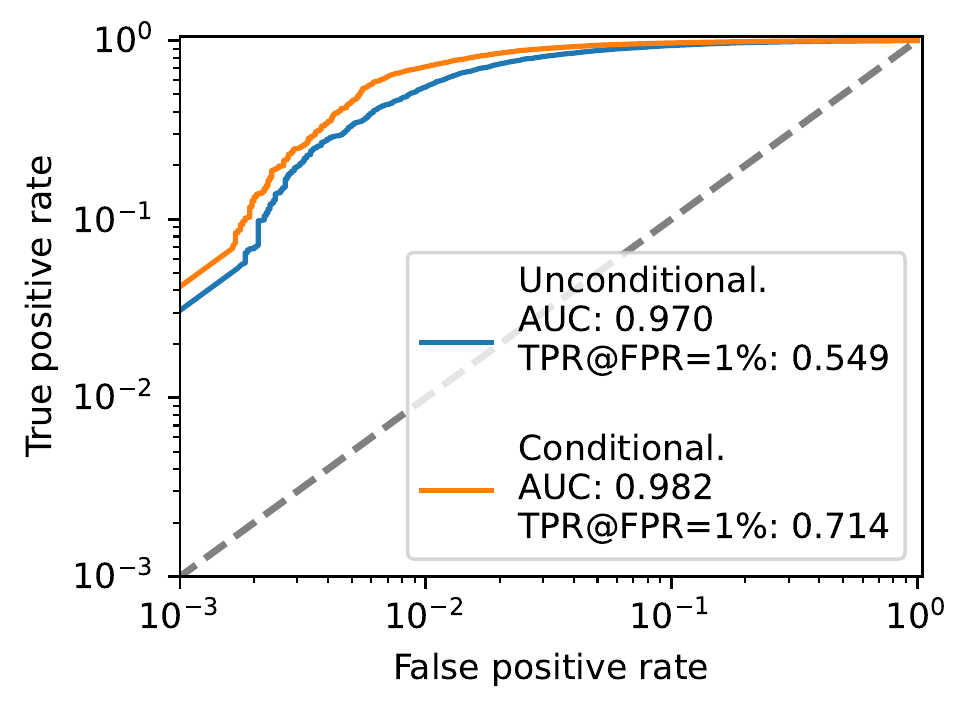}
        \caption{Membership attack against a conditional and unconditional diffusion model on CIFAR-10.}
         \label{fig:mia_cifar10_cond_vs_uncond}
\end{figure}

Diffusion models can be conditioned on labels (or prompts for text-to-image models).
We compare the difference in membership inference on a CIFAR-10 diffusion model trained unconditionally with a model conditionally trained on CIFAR-10 labels.
The conditional and unconditional models reach approximately the same FID after training; between 3.5-4.2 FID.
We plot the membership attack ROC curve in \Cref{fig:mia_cifar10_cond_vs_uncond} and note that the conditional model is marginally more vulnerable.
However, it is difficult to tell if this is a fundamental difference between conditional and unconditional models, or because the conditional model contains more parameters than unconditional model (the conditional models contains an extra embedding layer for the one-hot label input).

\newpage
\section{More Inpainting Attacks on CIFAR-10} \label{sec: cifar10_inpaint_more}

\begin{figure*}[h!]
\captionsetup{width=0.9\textwidth, justification=centering}
  \centering
\begin{subfigure}{0.33\textwidth}
\centering
    \includegraphics[width=0.95\linewidth]{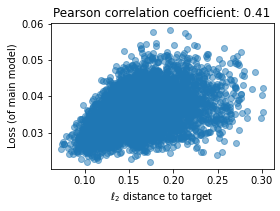}
        \caption{Loss (using the main model at diffusion timestep 100) on all 5,000 inpainted examples $X_{rec}$.}
    \label{fig:inpainting_rec_attack_train_bird_loss_points}
\end{subfigure}%
\begin{subfigure}{0.33\textwidth}
\centering
    \includegraphics[width=0.95\linewidth]{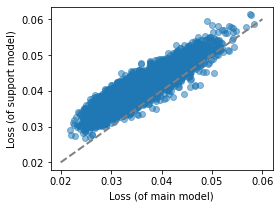}
        \caption{Comparison of loss on main and support models (at diffusion timestep 100) on all 5,000 inpainted examples.}
    \label{fig:train_bird_main_vs_support_loss}
\end{subfigure}%
\begin{subfigure}{0.33\textwidth}
\centering
    \includegraphics[width=0.95\linewidth]{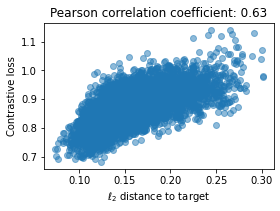}
        \caption{Contrastive loss ($\frac{\text{main model loss}}{\text{support model loss}}$) on all 5,000 inpainted examples $X_{rec}$.}
         \label{fig:inpainting_rec_attack_train_bird_contrast_points}
\end{subfigure}
\begin{subfigure}{0.45\textwidth}
\centering
    \includegraphics[width=0.95\linewidth]{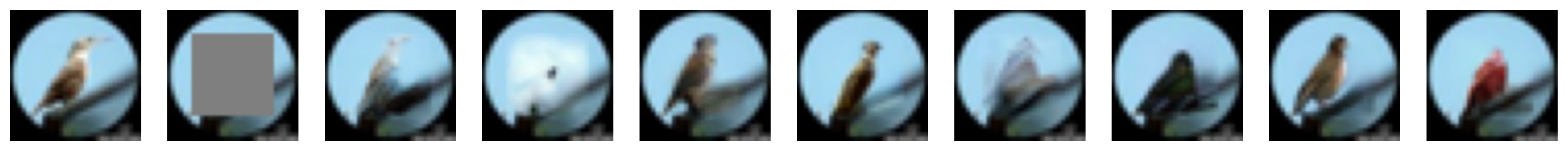}
        \caption{8 inpainted examples with the smallest loss. Leftmost is the original example, second to left is the masked example and the rest are inpainted examples.}
         \label{fig:inpainting_rec_attack_train_bird_loss_example}
\end{subfigure}%
\begin{subfigure}{0.45\textwidth}
\centering
    \includegraphics[width=0.95\linewidth]{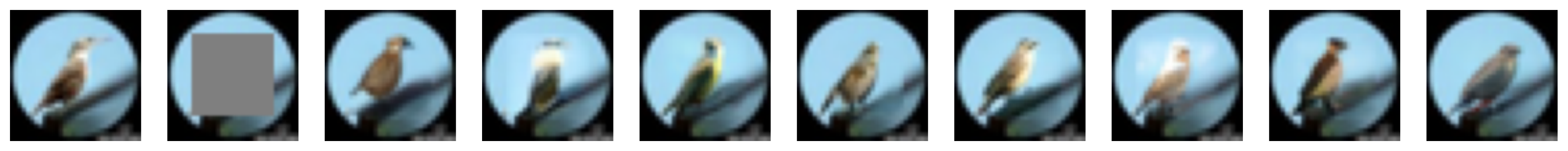}
        \caption{8 inpainted examples with the smallest $\frac{\text{main model loss}}{\text{support model loss}}$. Leftmost is the original example, second to left is the masked example and the rest are inpainted examples.}
         \label{fig:inpainting_rec_attack_train_bird_contrast_example}
\end{subfigure}%
\caption{Example of an inpainting attack (against a model we refer to as the \emph{main} model) on an image of a bird from CIFAR-10 when that image is included in training, and we mask out 60\% of the central pixels. In (a) we plot the $L_2$ distance between 5,000 inpainted reconstructions and the original (non-masked out) image and compare this to the loss with respect to the (main) model. In (b), we compare the loss of these reconstructions on the (main) model with a \emph{support model} for which we know the image wasn't contained in the training set. In (c), we compare $L_2$ distances between reconstructions with a contrastive loss which is given as the loss of the image with respect to the main model divided by the loss of the image with respect to the support model, and find there is stronger relationship between smaller $L_2$ distances and smaller losses compared to (a). Figure (d) gives examples of reconstructions with small loss and Figure (e) gives examples of reconstructions with small contrastive loss.}
\label{fig:inpainting_rec_attack_train_bird}
\end{figure*}

Here, we take a deeper dive into the inpainting attacks introduced in \Cref{ssec: cifar10_inpaint}.
As previously explained, for a target $x$, we create $X_{rec}$ where $|X_{rec}|=5000$.
In \Cref{fig:inpainting_rec_attack_train_bird_loss_points}, for every $x_{rec}\in X_{rec}$, we plot the normalized $\ell_2$ distance between the reconstruction and target, against the loss (at diffusion timestep 100) of $x_{rec}$.
We also plot in \Cref{fig:inpainting_rec_attack_train_bird_loss_example}, the eight examples from $X_{rec}$ that have the smallest loss on the main model.
There is a small positive correlation between loss and $\ell_2$ distance; although some appear to be similar to $x$, there are notable differences.

In \Cref{fig:train_bird_main_vs_support_loss} we compare the loss of each reconstruction on the main model against the \emph{support} model we will use to form the contrastive loss.
We make this correlation more pronounced by dividing the main loss by the support loss in  \Cref{fig:inpainting_rec_attack_train_bird_contrast_points}.
This has the effect of increasing the correlation between the (now contrastive) loss and $\ell_2$ distance.
This has the effect of filtering out examples that are seen as likely under both models, and can be seen by inspecting the eight examples from $X_{rec}$ that have have the smallest $\frac{\text{main model loss}}{\text{support model loss}}$ in \Cref{fig:inpainting_rec_attack_train_bird_contrast_example}.
These examples look more visually similar to $x$ in comparison to examples in \Cref{fig:inpainting_rec_attack_train_bird_loss_example}.

\begin{figure*}[h!]
\captionsetup{width=0.95\textwidth, justification=centering}
  \centering
\begin{subfigure}{0.33\textwidth}
\centering
    \includegraphics[width=0.95\linewidth]{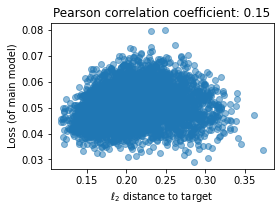}
        \caption{Loss (using the main model at diffusion timestep 100) on all 5,000 inpainted examples $X_{rec}$.}
    \label{fig:bird_mse_vs_rec}
\end{subfigure}%
\begin{subfigure}{0.33\textwidth}
\centering
    \includegraphics[width=0.95\linewidth]{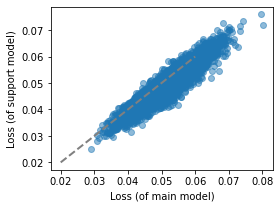}
        \caption{Comparison of loss on main and support models (at diffusion timestep 100) on all 5,000 inpainted examples.}
    \label{fig:test_bird_main_vs_support_loss}
\end{subfigure}%
\begin{subfigure}{0.33\textwidth}
\centering
    \includegraphics[width=0.95\linewidth]{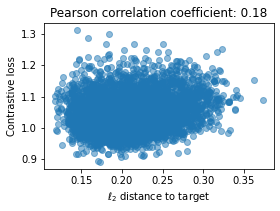}
        \caption{Contrastive loss ($\frac{\text{main model loss}}{\text{support model loss}}$) on all 5,000 inpainted examples $X_{rec}$.}
         \label{fig:bird_mse_vs_rec_contrast}
\end{subfigure}
\begin{subfigure}{0.45\textwidth}
\centering
    \includegraphics[width=0.95\linewidth]{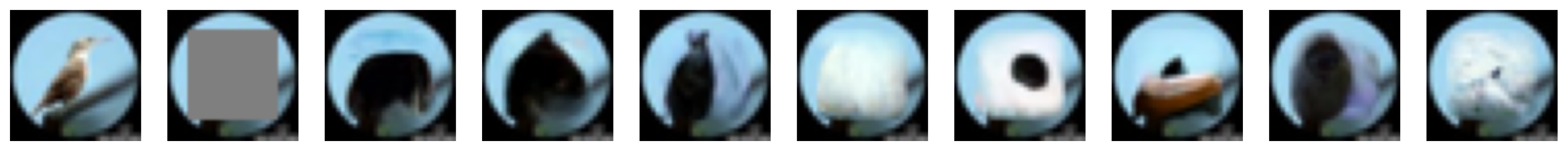}
        \caption{8 inpainted examples with the smallest loss. Leftmost is the original example, second to left is the masked example and the rest are inpainted examples.}
         \label{fig:bird_sort_by_loss}
\end{subfigure}%
\begin{subfigure}{0.45\textwidth}
\centering
    \includegraphics[width=0.95\linewidth]{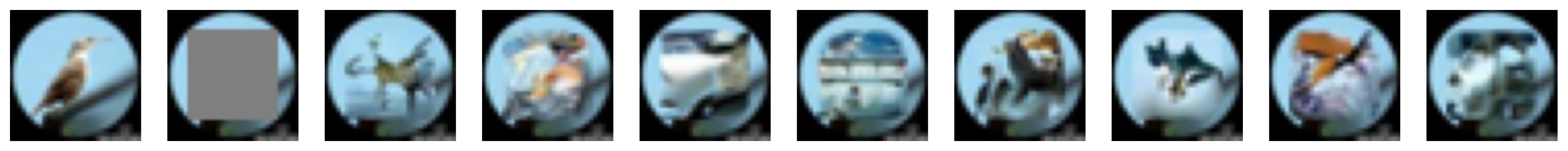}
        \caption{8 inpainted examples with the smallest $\frac{\text{main model loss}}{\text{support model loss}}$. Leftmost is the original example, second to left is the masked example and the rest are inpainted examples.}
         \label{fig:bird_sort_by_contrast_loss}
\end{subfigure}%
\caption{Example of an inpainting attack (against a model we refer to as the \emph{main} model) on an image of a bird from CIFAR-10 when that image is \emph{not} included in training, and we mask out 60\% of the central pixels. In (a) we plot the $L_2$ distance between 5,000 inpainted reconstructions and the original (non-masked out) image and compare this to the loss with respect to the (main) model. In (b), we compare the loss of these reconstructions on the (main) model with a \emph{support model} for which we know the image wasn't contained in the training set. In (c), we compare $L_2$ distances between reconstructions with a contrastive loss which is given as the loss of the image with respect to the main model divided by the loss of the image with respect to the support model, and find there is stronger relationship between smaller $L_2$ distances and smaller losses compared to (a). Figure (d) gives examples of reconstructions with small loss and Figure (e) gives examples of reconstructions with small contrastive loss.}
\label{fig:inpainting_rec_attack_test_bird}
\end{figure*}

\Cref{fig:inpainting_rec_attack_train_bird} inspected the attack success when $x$ was in the training set. 
We show in \Cref{fig:inpainting_rec_attack_test_bird} that the attack fails when $x$ was not included in training; using a contrastive loss doesn't signficantly increase the Pearson correlation coefficient. 
This means our attack is indeed exploiting the fact that the model can only inpaint correctly because of memorisation and not due to generalisation.
\newpage

\newpage

\section{GAN Training Setup}
We used on StudioGAN\footnote{\url{https://github.com/POSTECH-CVLab/PyTorch-StudioGAN}} codebase for training GAN in this work. For the StyleGAN and MHGAN architectures, we followed the default hyper-parameters provided in the StudioGAN repository. However, for the BigGAN architecture, we increased the number of training steps to 200,000, which is different from the original hyper-parameters, to increase image fidelity. We trained a total of 256 models for each GAN architecture, with each model being trained on a randomly selected half of the CIFAR-10 dataset. We selected the iteration that achieved the highest FID score on the test set for each model.

\section{Additional GAN Extraction Results}
\begin{figure*}[h!]
    \begin{subfigure}{\textwidth}
\centering
    \includegraphics[width=5.8in]{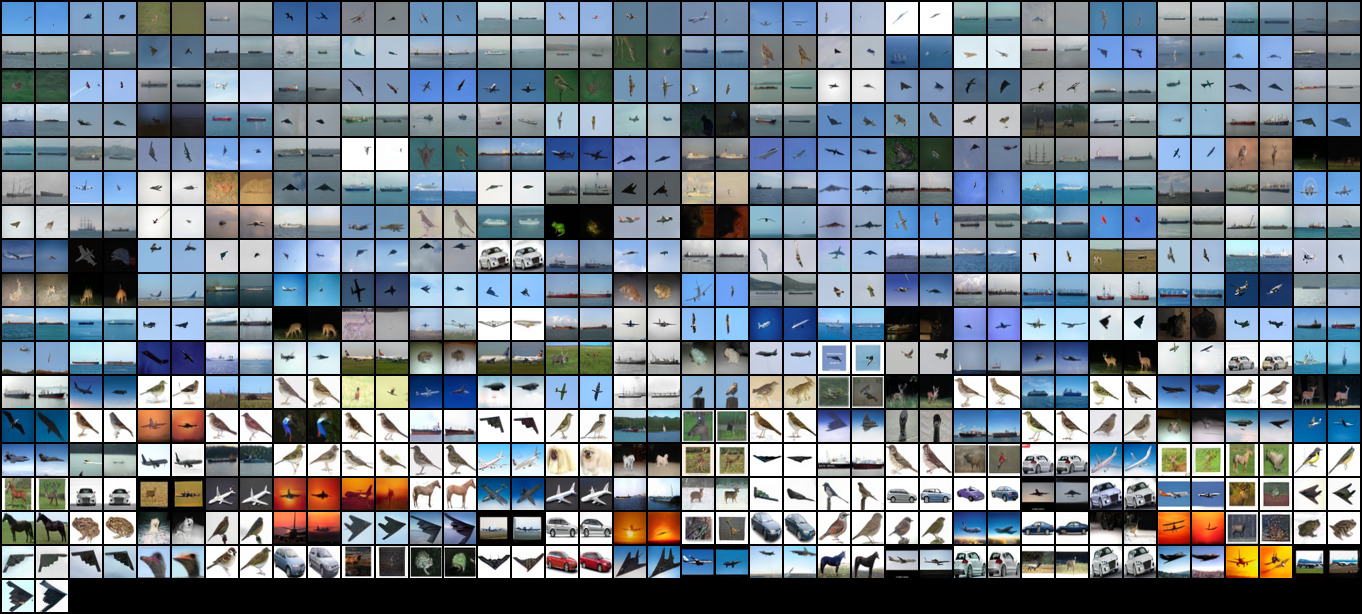}
        \caption{StyleGAN }
    \label{fig:stylegan_memorized}
\end{subfigure}%

\begin{subfigure}{\textwidth}
\centering
    \includegraphics[width=5.8in]{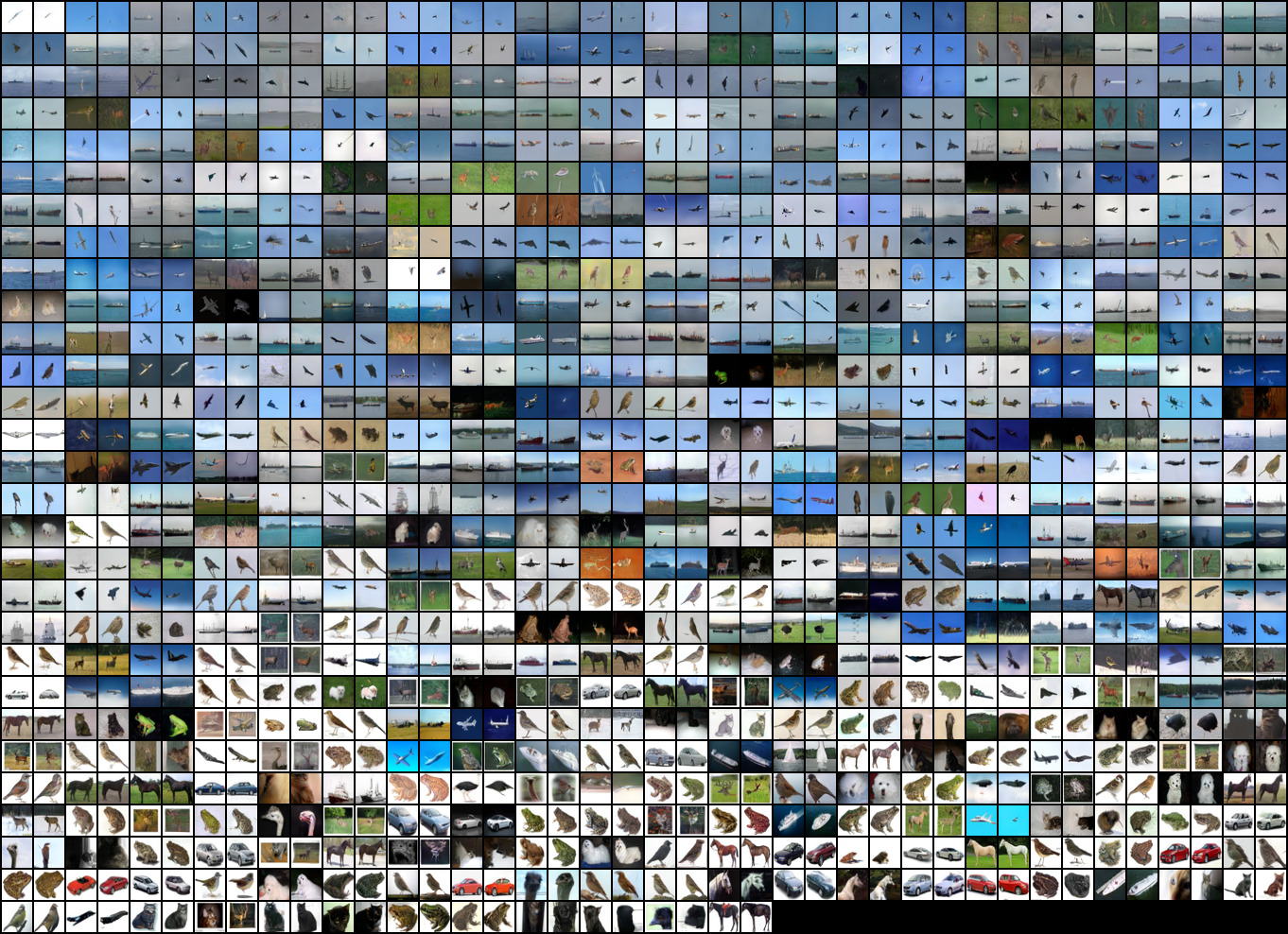}
        \caption{MHGAN }
    \label{fig:mhgan_memorized}
\end{subfigure}%

\begin{subfigure}{\textwidth}
\centering
    \includegraphics[width=5.8in]{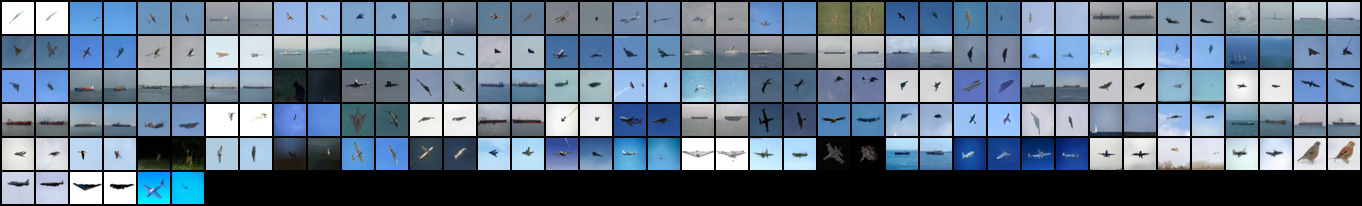}
        \caption{BigGAN }
    \label{fig:biggan_memorized}
\end{subfigure}%
    \caption{ Training examples extracted from a CIFAR-10 GAN for different architectures across $10^7$ generations.}
    \label{fig:gan_extractions_3}
\end{figure*}

\begin{figure*}[h!]
    \begin{subfigure}{\textwidth}
        \centering
        \includegraphics[width=\linewidth]{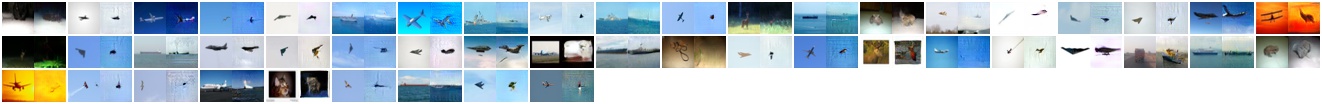}
            \caption{WGAN}
        \label{fig:wgan_normalized_l2}
    \end{subfigure}%
    
    \begin{subfigure}{\textwidth}
        \centering
        \includegraphics[width=\linewidth]{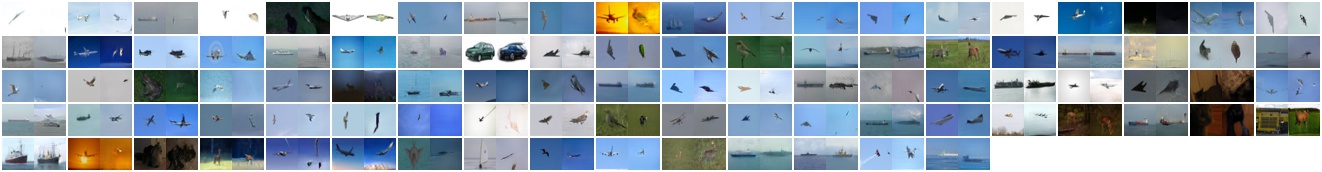}
            \caption{E2GAN}
        \label{fig:e2gan_normalized_l2}
    \end{subfigure}%
    
    \begin{subfigure}{\textwidth}
        \centering
        \includegraphics[width=\linewidth]{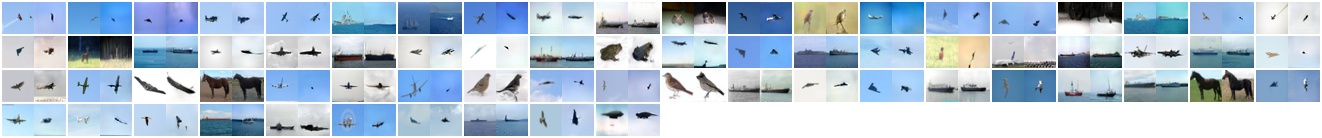}
            \caption{NDA}
        \label{fig:nda_normalized_l2}
    \end{subfigure}%
    
    \begin{subfigure}{\textwidth}
        \centering
        \includegraphics[width=\linewidth]{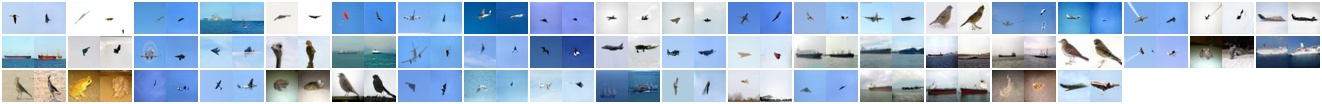}
            \caption{DiffAugment-BigGAN}
        \label{fig:diffbig_normalized_l2}
    \end{subfigure}%
    
    \begin{subfigure}{\textwidth}
        \centering
        \includegraphics[width=\linewidth]{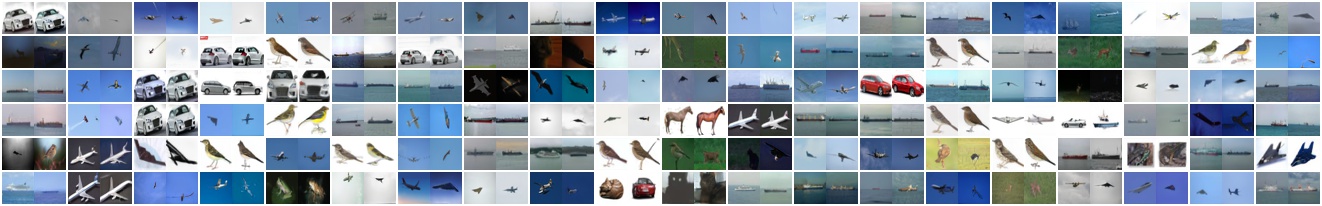}
            \caption{StyleGAN-ADA}
        \label{fig:stylegan_normalized_l2}
    \end{subfigure}%
    
    \begin{subfigure}{\textwidth}
        \centering
        \includegraphics[width=\linewidth]{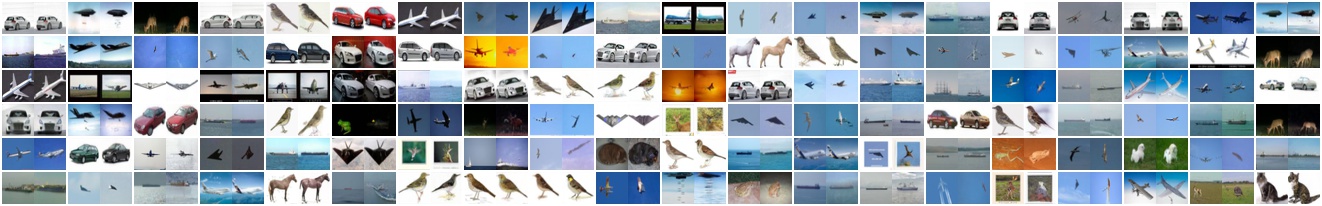}
            \caption{DDPM}
        \label{fig:ddpm_normalized_l2}
    \end{subfigure}%
    \caption{Training examples extracted from different publicly available pretrained GANs and diffusion (DDPM) models. We use normalized $\ell_2$ distance in pixel space to find memorized training samples. 
    In each pair of images, left and right image corresponds to real and it closely synthetic image.
    For StyleGAN-ADA and DDPM model we display 120 pairs with smallest normalized $\ell_2$ distance. For others we display all memorized training images.  1M generations
     }
    \label{fig:gan_cifar_all_extracted}
\end{figure*}

\Cref{fig:gan_extractions_3} and \Cref{fig:gan_cifar_all_extracted} contain additional examples extracted from GANs trained on CIFAR-10.

\end{document}